\documentclass[twocolumn,showpacs,amsmath,amssymb,superscriptaddress]{revtex4-1}
\usepackage{dcolumn}
\usepackage{color}
\usepackage{graphicx}
\usepackage{amsmath}
\usepackage{bm}
\usepackage{url}

\begin{document}

\title{Structural evolution in $A\approx 100$ nuclei within
the mapped 
interacting boson model based on the Gogny energy density
functional}

\author{K.~Nomura}
\affiliation{Physics Department, Faculty of Science, University of
Zagreb, HR-10000 Zagreb, Croatia}
\affiliation{Center for Computational
Sciences, University of Tsukuba, Tsukuba 305-8577, Japan}

\author{R.~Rodr\'iguez-Guzm\'an}
\affiliation{Physics Department, Kuwait University, 13060 Kuwait, Kuwait}

\author{L.~M.~Robledo}
\affiliation{Departamento de F\'\i sica Te\'orica, Universidad
Aut\'onoma de Madrid, E-28049 Madrid, Spain}

\date{\today}

\begin{abstract}
The structure of even-even neutron-rich Ru, Mo, Zr and Sr nuclei in the 
$A\approx 100$ mass region is studied within the interacting boson 
model (IBM) with microscopic input from the self-consistent mean-field 
approximation based on the Gogny-D1M energy density functional. The 
deformation energy surface in the quadrupole deformation space 
$(\beta,\gamma)$, computed  within the constrained  
Hartree-Fock-Bogoliubov framework, is mapped onto the expectation value 
of the appropriately chosen IBM Hamiltonian with configuration mixing 
in the boson condensate state. The mapped IBM Hamiltonian is used to 
study the spectroscopic properties of $^{98-114}$Ru, $^{96-112}$Mo, 
$^{94-110}$Zr and $^{92-108}$Sr. Several cases of $\gamma$-soft 
behavior are predicted in Ru and Mo nuclei while a pronounced 
coexistence between strongly-prolate and weakly-oblate deformed shapes 
is found for Zr and Sr nuclei. The method describes well the evolution 
of experimental yrast and non-yrast states as well as  selected $B$(E2) 
transition probabilities.
\end{abstract}



\maketitle


\section{Introduction}

The study of collective excitations and 
the associated shapes in nuclei 
with mass number $A \approx 100$ is 
receiving lately considerable attention in 
nuclear structure physics \cite{heyde11}. 
As a consequence of the subtle interplay between single-particle
and collective degrees of freedom, nuclei in this region of the nuclear chart
display a large variety of intriguing phenomena. Several experimental
\cite{sumikama2011,alb12,albers2013,thomas2013,browne2015,clement2016,park2016} and 
theoretical 
\cite{sieja2009,lalkovski2009,boyukata2010,rayner10odd-1,rayner10odd-2,mei2012,xiang2012,trodriguez2014,sarriguren2015}
studies have already been reported on the structure of those nuclei. In particular, the 
rapid structural change around the
neutron number $N=60$ has been carefully studied in 
Zr and Sr isotopes
\cite{federman1977,federman1978,skalski1997,rayner10odd-1,sumikama2011,mei2012,xiang2012,clement2016}. 
This mass region is also characterized by the competition between different low-lying configurations 
based on different intrinsic deformations, i.e., shape 
coexistence \cite{skalski1997,rayner10odd-1,park2016,mei2012,clement2016,heyde11}. 

Both the nuclear shell model (SM) \cite{caurier05} and the energy density functional 
(EDF) \cite{ben03rev} frameworks are among the most
popular theoretical tools used to describe the structure 
of medium-mass and heavy nuclei. On the one hand, the SM calculations  
encode the most important correlations for the 
considered nuclei and provide access to their 
spectroscopic properties. However, for open-shell systems, the 
dimension of the SM Hamiltonian matrix 
becomes exceedingly large  making its diagonalization 
prohibitively expensive. On the other hand, the EDF scheme 
\cite{ben03rev,Vre05,Nik11rev} yields a global
description of nuclear matter and bulk nuclear 
properties. Within 
this context, the evolution of the nuclear shapes 
around $A=100$ has been 
studied using self-consistent mean-field (SCMF) approximations based on  
different parametrizations of the  Skyrme \cite{bonche85} and Gogny \cite{rayner10odd-1,clement2016} 
as well as  relativistic \cite{lalazissis1995,xiang2012,mei2012} EDFs. However, in order
to access the excitation
spectra and transition rates within the EDF scheme, one needs to go beyond the mean-field level
to include dynamical correlations associated with the restoration of the broken 
symmetries  and/or the fluctuations in the 
collective coordinates. Such a task is usually accounted for via symmetry-projected 
configuration mixing calculations within the Generator Coordinate Method (GCM) 
framework \cite{RS,rayner02,ben03rev,Nik11rev}. However, symmetry-projected
GCM calculations are also computationally demanding in the case of 
heavy nuclear systems, especially when several collective coordinates (quadrupole, octupole, pairing, etc)
have to be taken into account as generating coordinates.

In this work, we resort to an alternative approach, which is based on
mapping the considered EDF into an algebraic model of interacting bosons \cite{nom08}. 
Our starting point is the (constrained) SCMF approximation that provides the 
corresponding microscopic energy surface as a function of the relevant deformation
parameters. Such a surface is subsequently mapped  onto
the  expectation value of the interacting boson model (IBM) \cite{IBM}
Hamiltonian computed with the boson condensate state. The parameters of the IBM Hamiltonian
are determined from such a mapping procedure. The resulting IBM Hamiltonian is then
used to obtain the  excitation spectra and
electromagnetic transition rates. The method allows a computationally feasible as well as quantitative   
description of the low-energy collective excitations. It has already
been applied  to  study  the
quadrupole \cite{nom08,nom10,nom11rot,nom12tri} and octupole \cite{nomura2015} modes in 
atomic  nuclei as well as to describe shape coexistence phenomena \cite{nom12sc,nom13hg,nomura2016}.
In the present study we extend the method of
Ref.~\cite{nom08} to study  the challenging structural evolution and  shape coexistence  in  neutron-rich 
nuclei with $A\approx 100$. 
The nuclei
$^{92-108}$Sr, $^{94-110}$Zr, $^{96-112}$Mo and $^{98-114}$Ru
have been taken as a representative sample. The phenomenological IBM framework has
been applied to describe some of those nuclei in the past 
\cite{isacker1980,sambataro82,garciaramos2005,sorgunlu2008,lalkovski2009,boyukata2010}. 
However, even when several attempts have been made 
\cite{lalkovski2009,boyukata2010,albers2013} to extrapolate the IBM scheme to unknown 
regions of the periodic table, in most of the cases the model has not been extensively
used to predict the properties of exotic nuclei. Within this 
context, one of the main advantages of our method over the 
conventional phenomenological IBM approaches is that, it predicts the  properties
of unexplored nuclei based only on the  underlying microscopic EDF framework. 
Our SCMF calculations are based on the Gogny-D1M \cite{Gogny,D1M} EDF. Previous 
studies \cite{rayner10pt,rayner12,robledo13,rayner2014,robledo2015} have shown, that the 
parameter set D1M essentially keeps the same predictive 
power of the more traditional Gogny-D1S \cite{D1S}  EDF 
regarding  a wealth of low-energy nuclear structure properties while 
improving the description of the nuclear masses \cite{D1M}. The Gogny-D1M
EDF has also been applied to the study of nuclei, including the odd-mass ones,
in the $A\approx 100$ mass region \cite{rayner10odd-1,rayner10odd-2,alb12,albers2013}.

The paper is organized as follows. In Sec.~\ref{sec:theory} we briefly 
outline the theoretical framework used in this work. The results of our 
calculations are presented  in Sec.~\ref{sec:results} where, we discuss 
the deformation energy surfaces and the IBM parameters derived from our 
mapping procedure. In the same section we consider the evolution of the 
low-lying levels in the studied nuclei as well as the systematics of 
the $B$(E2) transition rates. We also discuss the individual level 
schemes for $N=60$ isotones. The robustness of our method is addressed
by studying specifically the sensitivity of the results to variations on key
parameters. Finally, Sec.~\ref{sec:summary} is devoted  
to the concluding remarks.


\section{Theoretical  framework \label{sec:theory}}


\subsection{Constrained SCMF calculations}

As already mentioned above, the first step in our calculations is to 
obtain the deformation energy surfaces for the considered nuclei. To 
this end, we have performed constrained Hartree-Fock-Bogoliubov (HFB) 
calculations  based on the Gogny-D1M EDF. We have  resorted to 
constrains on the $\hat{Q}_{20}$ and $\hat{Q}_{22}$ operators 
\cite{rayner10pt,robledo2008}. The quadrupole moment is then defined as 
$Q=\sqrt{Q_{20}+Q_{22}}$. We then consider the deformation parameters 
$\beta=\sqrt{4\pi/5}Q/\langle r^2\rangle$ and 
$\gamma=\tan^{-1}{Q_{22}/Q_{20}}$. In the definition of $\beta$ the 
mean-square radius $\langle r^2\rangle$ is evaluated with the 
corresponding HFB state. For more details on the constrained Gogny-HFB 
framework the reader is referred, for example, to 
Ref.~\cite{rayner10pt}. In what follows,  we will refer to the total 
mean-field energy  as a function of the $(\beta,\gamma)$ parameters as 
the deformation energy surface.


\subsection{IBM with configuration mixing}

In order to compute the  excitation spectra and transition rates,  we 
use the method of Ref.~\cite{nom08} in which the parameters of the IBM
Hamiltonian are determined by mapping the 
deformation energy surface provided 
by the constrained Gogny-D1M SCMF calculations onto the expectation 
value of the IBM Hamiltonian computed the boson condensate (intrinsic) wave function
\cite{GK}. The resulting IBM Hamiltonian is then used to calculate spectroscopic 
properties for the studied nuclei. We have considered the 
proton-neutron IBM (denoted IBM-2) \cite{OAIT,OAI} as it represents a more realistic 
approach, able to treat  both the proton and neutron degrees of freedom. 
The building blocks of the  IBM-2 model are  the correlated 
monopole 0$^+$ ($S_{\pi}$ and $S_{\nu}$) and 
quadrupole $2^+$
($D_{\pi}$ and $D_{\nu}$)
pairs of valence protons ($\pi$) and
neutrons ($\nu$). The $S_{\pi}$ ($S_{\nu}$) and $D_{\pi}$ ($D_{\nu}$) pairs are associated with
the proton (neutron) $s_{\pi}$ ($s_{\nu}$) and $d_{\pi}$ ($d_{\nu}$)
bosons, which have spin and parity $J^{\pi}=0^+$ and $2^+$, respectively \cite{OAI}. 
The number of proton ($N_{\pi}$) and neutron ($N_{\nu}$) bosons is equal to
half the number of valence protons and neutrons \cite{OAIT,OAI}. The bosonic model
space comprises the neutron major shell
$N=50-82$ and the proton $Z=40-50$ shell for Ru, Mo and Zr isotopes and
$Z=28-40$ in the case of Sr isotopes. Therefore,  2 $\le$ $N_{\nu}$ $\le$ 8  for the studied nuclei
while   $N_{\pi}=0$
(Zr), 1 (Sr and Mo) and 2 (Ru).

As will be shown, for many of the  nuclei in the selected sample, the Gogny-D1M  energy
surface exhibits up to three mean-field minima close in energy to each
other. 
Accordingly, the bosonic  model space should be  extended so as to take into account those
configurations. 
In a mean-field picture, the different mean-field minima are associated
with  $2n$-particle-$2n$-hole ($n=0,1,2$) intruder excitations
across the closed shell. 
To incorporate the intruder configurations, we follow the method of
Duval and Barrett \cite{duval81} which associates the different SM-like spaces of $0p-0h$,
$2p-2h$, $4p-4h$, \ldots 
excitations with the corresponding boson spaces comprising  $N_B$,
$N_B+2$, $N_B+4$, \ldots 
bosons, where $N_B(=N_\nu + N_{\pi})$ denotes the total number of bosons.
The different boson subspaces are allowed to mix by introducing an additional interaction. 
Under the assumption of  Duval and Barrett, particles and holes are not
distinguished. 
Then, as the excitation of one pair (boson) increases the boson number
by 2, the configurations for the $2np-2nh$ excitations differ
from each other in boson number by 2. 
In the following, we assume only the proton $ph$ excitations across the sub-shell
closure $Z=40$, which is equivalent to the excitation from the proton
$pf$ shell to the $1g_{9/2}$ orbital. 
The Hilbert space for the configuration mixing IBM-2 model  is then defined as the direct
sum of each ``unperturbed'' configuration
space, i.e., 
\begin{eqnarray}
 [N_{\nu}\otimes N_{\pi}]\oplus [N_{\nu}\otimes
(N_{\pi}+2)]\oplus [N_{\nu}\otimes (N_{\pi}+4)], 
\end{eqnarray}
where $[N_{\nu}\otimes (N_{\pi}+2n)]$ ($n=0$, 1 and 2) denotes the 
configuration space for the unperturbed IBM-2 Hamiltonian for the
$2np-2nh$ proton excitations, comprising 
$N_{\nu}$ neutron and $N_{\pi}+2n$ proton bosons. 
In the following, the unperturbed space $[N_{\nu}\otimes (N_{\pi}+2n)]$
is simply denoted  as $[n]$ ($n=0$, 1 and 2), and we refer to the IBM-2
simply as IBM, unless otherwise specified.

The Hamiltonian $\hat H_B$ for the system is then expressed in terms of 
up to three unperturbed IBM Hamiltonians $\hat H_n$ ($n=0$, 1 and 2) differing in
boson number by 2 and in  terms of $\hat H_{n,n+1}^{mix}$ that mix
different boson subspaces: 
\begin{eqnarray}
\label{eq:ham-cm}
 \hat H_B = \hat H_{0} + (\hat H_{1}+\Delta_{1}) + (\hat
  H_{2}+\Delta_{2})+\hat
  H^{mix}_{0,1} + \hat H^{mix}_{1,2}, 
\end{eqnarray}
where $\Delta_{1}$ and $\Delta_{2}$ represent the
energies  required to excite one and two bosons across the inert core.

For the unperturbed Hamiltonian $\hat H_{n}$ ($n=0$, 1 and 2) we  have taken the  form 
\begin{eqnarray}
\label{eq:ham-sg}
 \hat H_{n} = \epsilon_n\hat n_d + \kappa_n\hat Q\cdot\hat Q +
  \kappa^{\prime}_n\sum_{\rho^{\prime}\neq\rho}\hat T_{\rho\rho\rho^{\prime}}. 
\end{eqnarray}
where the first term $\hat n_d=\hat n_{d\nu} + \hat n_{d\pi}$, with $\hat
n_{d\rho}=d^{\dagger}_{\rho}\cdot\tilde d_{\rho}$ ($\rho=\nu,\pi$), represents the
$d$-boson
number operator. On the other hand, $\hat Q = \hat Q_{\nu} + \hat Q_{\pi}$ is the quadrupole operator and
$\hat Q_{\rho}=s^{\dagger}_{\rho}\tilde
d_{\rho}+d^{\dagger}_{\rho}\tilde
s_{\rho}+\chi_{\rho,n}[d^{\dagger}_{\rho}\times\tilde d_{\rho}]^{(2)}$. 
The third term is a specific three-boson interaction with $\hat T_{\rho\rho\rho^{\prime}} =
\sum_{L}[d^{\dagger}_{\rho}\times d^{\dagger}_{\rho}\times
d^{\dagger}_{\rho^{\prime}}]^{(L)}\cdot [\tilde d_{\rho^{\prime}}\times
\tilde d_{\rho}\times\tilde d_{\rho}]^{(L)}$, where $L$ denotes the
total angular momentum in the boson system. 
 
The mixing interaction reads 
\begin{eqnarray}
\label{eq:ham-mix}
 \hat H^{mix}_{n,n+1} = \omega_{s,n} s^{\dagger}_{\pi}\cdot s^{\dagger}_{\pi}+\omega_{d,n}
  d^{\dagger}_{\pi}\cdot d^{\dagger}_{\pi} + h.c.
\end{eqnarray}
where $\omega_{s,n}$ and $\omega_{d,n}$ are strength parameters,  assumed
to be equal $\omega_{s,n}=\omega_{d,n}=\omega_n$, for simplicity.  
Note that there is no direct coupling between the $[n=0]$ and $[n=2]$
spaces with the two-body nuclear interactions.

The unperturbed Hamiltonian $\hat H_{n}$ in Eq.~(\ref{eq:ham-sg}) takes the 
simplest form of the $sd$-IBM-2 Hamiltonian used for describing low-energy quadrupole
collective states. It is only composed of $\hat n_d$ 
and $\hat Q_{\nu}\cdot\hat Q_{\pi}$ terms \cite{OAI,mizusaki1996,nom08}. 
The addition of the like neutron boson term $\hat Q_{\nu}\cdot\hat Q_{\nu}$ in
Eq.~(\ref{eq:ham-sg}) is due to the fact that, in the present study,
$N_{\pi}=0$ for the normal (or $0p-0h$) configuration of the
Zr isotopes and without this term the SCMF minimum could not be reproduced.
We also include the interaction term between like proton bosons $\hat
Q_{\pi}\cdot\hat Q_{\pi}$ and, to reduce the number of parameters, assume
the $F$-spin \cite{OAIT,iachello1984} invariant form for the quadrupole operator $\hat Q=\hat Q_{\nu} + \hat
Q_{\pi}$. On the other hand, the three-boson term is required to describe a 
triaxial minimum. We only consider the interaction between proton and
neutron bosons  with $L=3$. The specific choice of the three-boson term is due to the 
relevance of the proton-neutron interactions 
in  medium-mass and heavy nuclei, 
and that only the $L=3$ term gives rise to a stable triaxial minimum at
$\gamma\approx 30^{\circ}$ \cite{nom12tri}. 
For those
nuclei where the configuration mixing is taken into account, the strength
parameter $\kappa^{\prime}$ is taken to be equal to that of the
quadrupole-quadrupole term, i.e., $\kappa^{\prime}=\kappa$.  
For the nuclei where the configuration mixing is not
considered, $\kappa^{\prime}$ is taken as an independent parameter.

To look at the geometrical feature of the configuration-mixing IBM Hamiltonian $\hat H_B$, we 
introduce the following boson intrinsic state 
$|\Phi_B(\beta,\gamma)\rangle$, extended to the space $[n=0]\oplus
[n=1]\oplus [n=2]$: 
\begin{eqnarray}
\label{eq:coherent-cm}
 |\Phi_B(\beta,\gamma)\rangle=|\Phi_B(0,\beta,\gamma)\rangle\oplus
  |\Phi_B(1,\beta,\gamma)\rangle\oplus |\Phi_B(2,\beta,\gamma)\rangle. 
\nonumber \\
\end{eqnarray}
The coherent state for each unperturbed space $|\Phi_B(n,\beta,\gamma)\rangle$ ($n=0$, 1 and 2) reads 
\begin{eqnarray}
 |\Phi(n,\beta,\gamma)\rangle=\frac{1}{\sqrt{N_{\nu}!N_{\pi,n}!}}(\lambda_{\nu}^{\dagger})^{N_{\nu}}(\lambda_{\pi}^{\dagger})^{N_{\pi,n}}|0\rangle, 
\end{eqnarray}
with $N_{\pi,n}\equiv N_{\pi}+2n$ and 
\begin{eqnarray} \lambda_{\rho}=s^{\dagger}_{\rho}+\beta_{\rho}\cos{\gamma_{\rho}}d_0^{\dagger}+\frac{1}{2}\beta_{\rho}\sin{\gamma_{\rho}}(d^{\dagger}_{+2}+d^{\dagger}_{-2}). 
\end{eqnarray}
$\beta_{\rho}$ and $\gamma_{\rho}$ are the quadrupole deformation
parameters analogous to those in the collective model \cite{BM}. They are assumed to be the same between
protons and neutrons, i.e., $\beta_{\nu}=\beta_{\pi}\equiv\beta_B$ and
$\gamma_{\nu}=\gamma_{\pi}\equiv\gamma_B$. 
The bosonic deformation parameters $\beta_{B}$ and $\gamma_B$ could be
related to those in the collective model in such a way that 
$\beta_B\propto\beta$ and $\gamma_B=\gamma$  \cite{GK}. 

The expectation value of the total Hamiltonian $\hat H_B$ in the  
coherent state $|\Phi_B(\beta,\gamma)\rangle$ leads to consider the $3\times 3$ matrix
\cite{frank04}: 
\begin{eqnarray}
\label{eq:pes}
  {\cal E}=\left(
\begin{array}{ccc}
E_{0}(\beta,\gamma) & \Omega_{0,1}(\beta) & 0 \\
\Omega_{1,0}(\beta) & E_{1}(\beta,\gamma)+\Delta_{1} & \Omega_{1,2}(\beta)
 \\
0 & \Omega_{2,1}(\beta) & E_{2}(\beta,\gamma)+\Delta_{2} \\
\end{array}
\right), 
\end{eqnarray}
where the diagonal and off-diagonal elements  represent
the expectation values of the unperturbed Hamiltonians and 
the mixing interactions, respectively.
The three eigenvalues of ${\cal E}$ correspond to specific energy
surfaces depending on the values of the parameters and it is customary to
take the lowest-energy one  \cite{frank04} as the IBM deformation energy 
at each deformation $(\beta,\gamma)$.
 
The analytical expression of the diagonal matrix
element $E_n(\beta,\gamma)$ ($n=0$, 1 and 2) is given as
\begin{eqnarray}
\label{eq:pes-detail1}
E_{n}(\beta,\gamma)=&&\frac{k_1+k_2\beta_{B,n}^2}{1+\beta_{B,n}^2}
+\frac{k_3\beta_{B,n}^2+k_4\beta_{B,n}^3\cos{3\gamma}+k_5\beta_{B,n}^4}{(1+\beta_{B,n}^2)^2}
\nonumber \\
&&+\frac{k_6\beta_{B,n}^3\sin^2{3\gamma}}{(1+\beta_{B,n}^2)^3}
\end{eqnarray}
where 
\begin{eqnarray}
\label{eq:pes-detail1-1}
&&k_1=5\kappa_n(N_{\nu}+N_{\pi.n}) \nonumber \\
&&k_2=[\epsilon_n+\kappa_n(1+\chi^2_{\nu,n})]N_{\nu}+[\epsilon_n+\kappa_n(1+\chi^2_{\pi,n})]N_{\pi,n}
 \nonumber \\
&&k_3=4\kappa_n(N_{\nu}+N_{\pi,n})(N_{\nu}+N_{\pi,n}-1) \nonumber \\
&&k_4=-4\kappa_n\sqrt{\frac{2}{7}}(\chi_{\nu,n}N_{\nu}+\chi_{\pi,n}N_{\pi,n})(N_{\nu}+N_{\pi,n}-1)
 \nonumber \\
&&k_5=\frac{2}{7}\kappa_n[(\chi_{\nu,n}N_{\nu}+\chi_{\pi,n}N_{\pi,n})^2-(\chi_{\nu,n}^2N_{\nu}+\chi_{\pi,n}^2N_{\pi,n})]
 \nonumber \\
&&k_6=-\frac{1}{7}\kappa^{\prime}_nN_{\nu}N_{\pi,n}(N_{\nu}+N_{\pi,n}-2), 
\end{eqnarray}
and that of the non-diagonal matrix element $\Omega_{n,n+1}(\beta)$
($n=0$ and 1) as
\begin{eqnarray}
\label{eq:pes-detail2}
\Omega_{n,n+1}(\beta)&=&\Omega_{n+1,n}(\beta)
\nonumber \\
&=&\omega_n\sqrt{(N_{\pi,n}+1)N_{\pi,n+1}}\times 
\nonumber \\
&&
\Big[\frac{1+\beta_{B,n}\beta_{B,n+1}}{\sqrt{(1+\beta_{B,n}^2)(1+\beta_{B,n+1}^2)}}\Big]^{N_{\nu}+N_{\pi,n}}. 
\end{eqnarray}
In both Eq.~(\ref{eq:pes-detail1}) and (\ref{eq:pes-detail2})
$\beta_{B,n}$ denotes the bosonic deformation parameter in each 
unperturbed space 
$[n]$, and is connected to the  $\beta$ deformation parameter of the SCMF model through
the relation $\beta_{B,n}=C_{n}\beta$.  The constant 
$C_n$ is also determined from the energy-surface fitting procedure
by requiring that the position of the minimum for each unperturbed configuration is reproduced. 
The formulas in Eqs.~(\ref{eq:pes-detail1}) and (\ref{eq:pes-detail2}) 
are the same as those found
in Ref.~\cite{nom13hg} except for the fact that, in the present study,
the expectation value of the like-particle $\hat
Q_{\rho}\cdot\hat Q_{\rho}$ term is also included while that of the rotational $\hat
L\cdot\hat L$ term is not included.


\subsection{The mapping procedure}

All together the model has 22 parameters that have to be fixed. This
represents too much freedom and therefore some of the parameters have
been kept fixed to simplify the calculation. The fitting protocol used
is the following 
  
\begin{enumerate}
 \item[(i)] Each unperturbed Hamiltonian is determined by using
	    the procedure of Refs.~\cite{nom08,nom10}: 
	    each diagonal matrix element $E_{n}$ in Eq.~(\ref{eq:pes})
	    is fitted to the corresponding mean-field minimum. The
	    normal $[n=0]$ configuration is assigned to the mean-field minimum
	    with the smallest deformation \cite{naza93}. On the 
	    other hand the $[n=1]$ ($[n=2]$) configuration
	    corresponds to
	    the minimum with the second (third) smallest deformation. In this way, each 
	    unperturbed Hamiltonian is determined
	    independently.	   	   	    
\item[(ii)] We then
	    extract the energy offsets $\Delta_{1}$ and
	    $\Delta_{2}$ so that the energy difference between the
	    two neighboring minima on the Gogny-D1M energy surface is
	    reproduced. 
\item[(iii)] Finally, what is left is to introduce the mixing interactions $\hat 
	     H^{mix}_{n,n+1}$ and determine the $\omega$ strengths.
	     They could be determined so as to reproduce the
	     topology of the barriers between  the minima Refs.~\cite{nom12sc,nom13hg}.
	     However, in this work we have assumed, for the sake of simplicity, a constant strength
	     $\omega=0.1$ MeV for both $\hat H^{mix}_{0,1}$ and $\hat
	     H^{mix}_{1,2}$ terms, in order to 
	      keep the mixing interactions perturbative. 
\end{enumerate}

Once its parameters are determined, the Hamiltonian $\hat H_B$ is diagonalized 
in the $[n=0]\oplus [n=1]\oplus [n=2]$  space using the boson $m$-scheme 
\cite{nomura-phdthesis}. The resulting wave function is then used to compute
the electromagnetic E2 transition rates. The E2 transition operator is given as 
\begin{eqnarray}
 \hat T^{(E2)} = \sum_{n=0,1,2}e_{B,n}\hat Q_n, 
\end{eqnarray}
where $e_{B,n}$ and $\hat Q_{n}$ are the effective charge and
quadrupole operator for the configuration $[n]$, respectively. For
simplicity, the effective charges are assumed to be the same for 
the three configurations, i.e.,  $e_{B,n=0}=e_{B,n=1}=e_{B,n=2}$. They are
fitted to reproduce the available experimental $B$(E2; $2^+_1\rightarrow
0^+_1$) values for the $N=66$  Ru, Mo and Zr nuclei
while for the Sr isotopes they are fitted to reproduce the 
experimental $B$(E2; $2^+_1\rightarrow
0^+_1$) value for  $^{100}$Sr. 


\section{Results and discussion\label{sec:results}}


\subsection{The Gogny-D1M  energy surfaces}

\begin{figure*}[htb!]
\begin{center}
\includegraphics[width=0.8\linewidth]{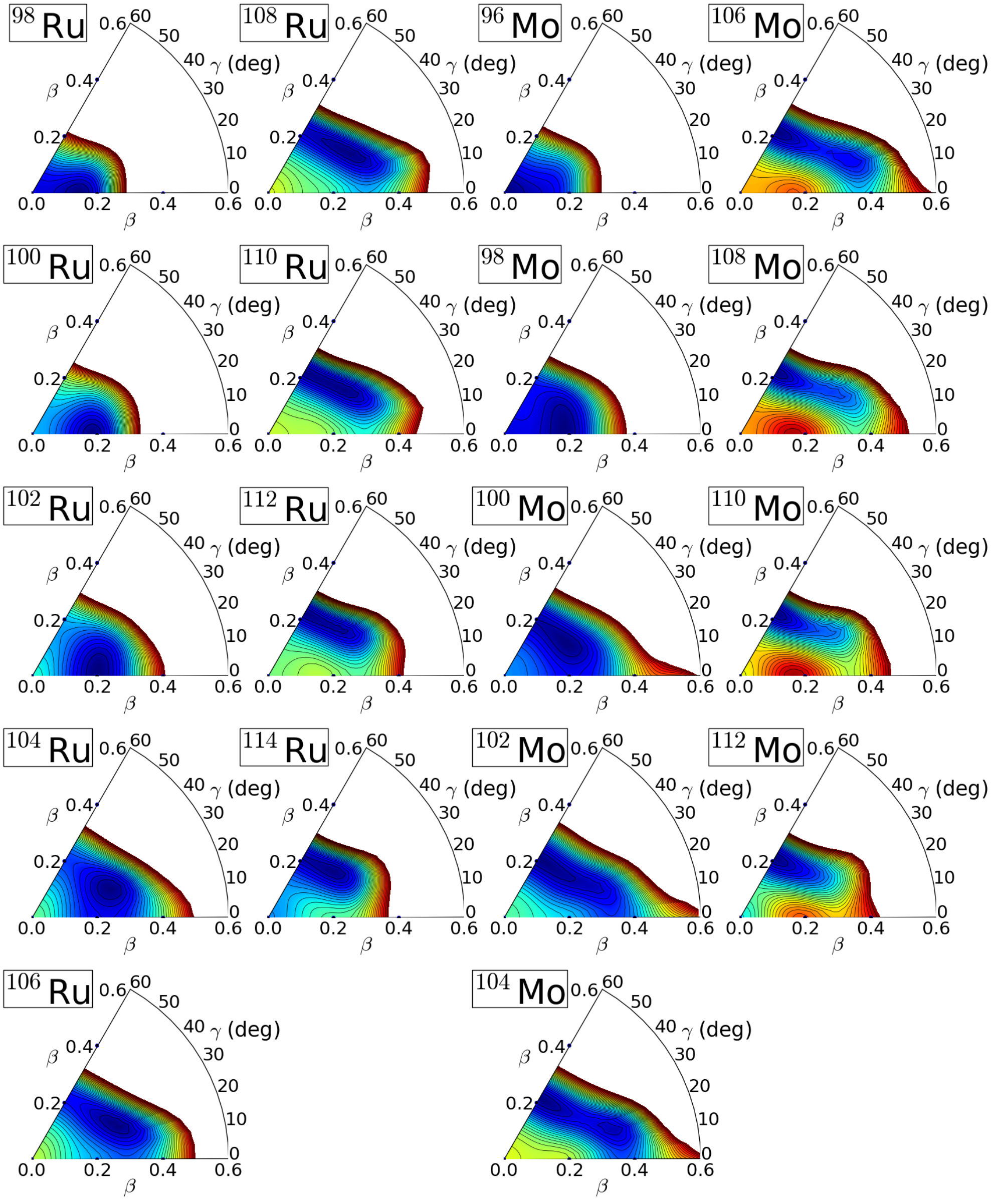}
\caption{(Color online) Contour plot of the deformation energy surfaces
in the $(\beta,\gamma)$ plane for the considered 
Ru and Mo isotopes from neutron number $N=54$ to 70, computed
with the constrained HFB method using the Gogny functional D1M. They
are plotted in the range $0.0\leq\beta\leq 0.6$ and
$0^{\circ}\leq\gamma\leq 60^{\circ}$. The difference between the
neighboring contours is 100 keV. }
\label{fig:pes-hfb-rumo}
\end{center}
\end{figure*}

\begin{figure*}[htb!]
\begin{center}
\includegraphics[width=0.8\linewidth]{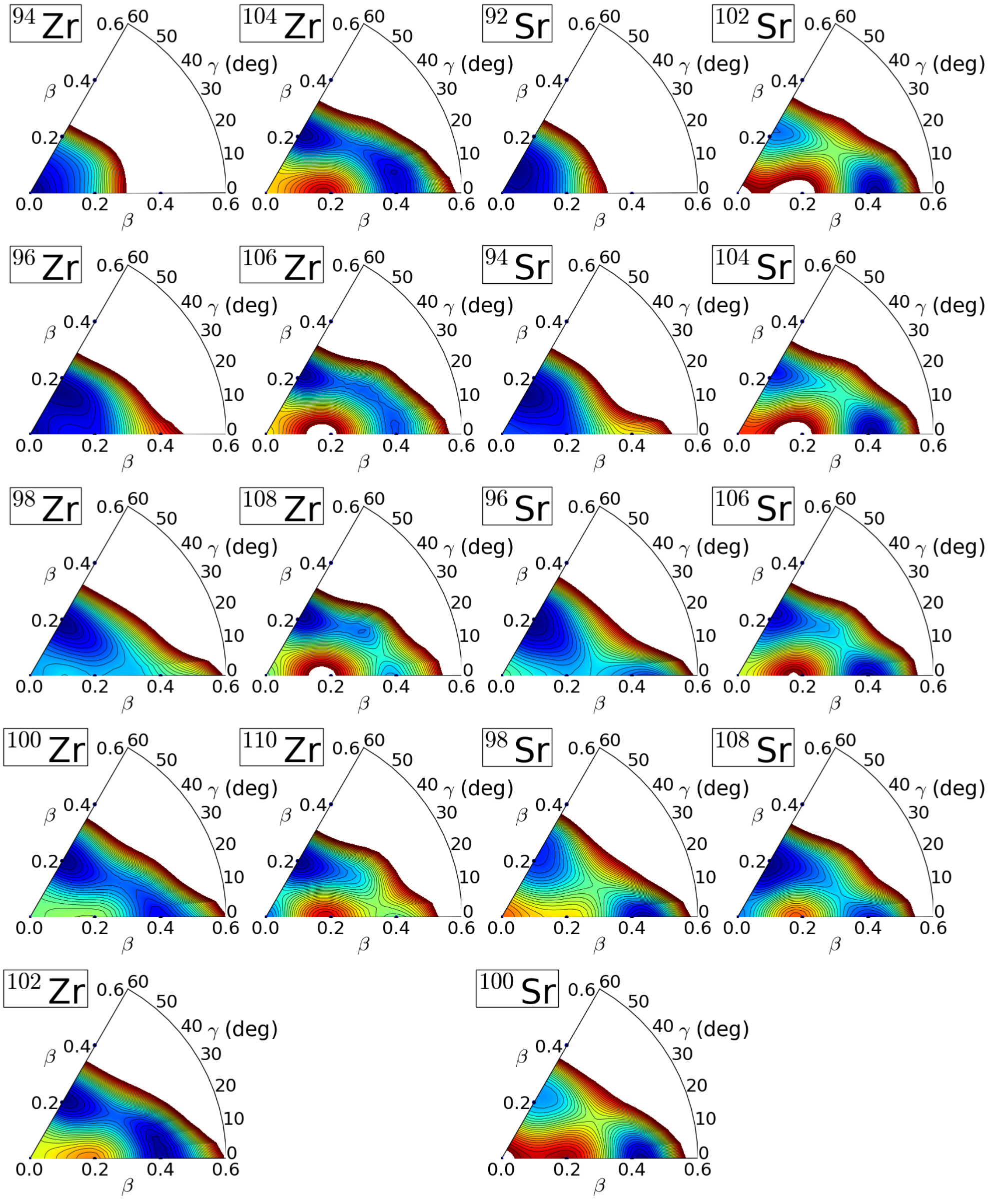}
\caption{(Color online) The same as in 
 Fig.~\ref{fig:pes-hfb-rumo}, but for the Zr and Sr isotopes. }
\label{fig:pes-hfb-zrsr}
\end{center}
\end{figure*}

In this section, we discuss the results of our SCMF calculations. In 
Figs.~\ref{fig:pes-hfb-rumo} and \ref{fig:pes-hfb-zrsr} we have 
depicted the deformation energy surfaces, obtained within the  
constrained Gogny-D1M EDF framework, for the considered Ru, Mo, Zr and 
Sr even-even nuclei with neutron numbers  54 $\le N\le$ 70. 

As can be seen from Fig.~\ref{fig:pes-hfb-rumo}, the lightest of the 
considered Ru isotopes exhibits a weakly deformed minimum. On the other 
hand, for $N$=60 the ground state corresponds to a shallow triaxial 
configuration around $\gamma=30^{\circ}$. In fact, the nucleus 
$^{104}$Ru is the softest in the $\gamma$ direction among all the Ru 
isotopes shown in the figure. Moreover, the ground state minimum 
remains triaxial up to $N$=68. For larger neutron numbers, the ground 
state  becomes oblate though it still remains $\gamma$-soft. For the 
studied Ru nuclei, the mean-field energy surfaces do not display 
multiple minima.

In the case of the Mo isotopes, also shown in  
Fig.~\ref{fig:pes-hfb-rumo}, the energy surface corresponding to 
$^{96}$Mo displays a nearly spherical minimum while for increasing 
neutron number the  surfaces become $\gamma$-soft up to $N$=62. 
Previous Skyrme Hartree-Fock plus BCS calculations \cite{thomas2013}, 
based on the SLy6 parameter set \cite{chabanat1997}, predicted two 
minima, one nearly spherical and the other triaxial, for $^{98}$Mo. 
On the other hand, in our Gogny-D1M SCMF calculations no coexisting minima 
are found for $^{98}$Mo, as well as for $^{100,102}$Mo. 
Two minima are found in the SCMF energy surfaces from around 
$N$=62, one oblate and the other triaxial with $\gamma$ around 
$20^{\circ}-30^{\circ}$. 
However, the heavier isotopes are less 
$\gamma$-soft with coexisting oblate and nearly spherical 
configurations in the case of $^{112}$Mo.

The systematics of the HFB energy surfaces, depicted in 
Fig.~\ref{fig:pes-hfb-zrsr} for the Zr and Sr isotopes, reveals a 
pronounced competition between oblate and prolate configurations. In 
the case of Zr nuclei, the oblate minimum remains the ground state up 
to $^{100}$Zr. The two mean-field minima found for $N=62$ and 64 are 
quite close in energy whereas the global minimum is still found on the
oblate side.  
For $N\geq 66$, the prolate minimum becomes less pronounced. 
For the Sr isotopes, a clear prolate minimum is found for $N$=60, 62 
and 64.  The energy surfaces obtained for $^{104,106}$Sr display two 
almost degenerate minima while the  oblate one becomes more pronounced 
for $^{108}$Sr. The previous results agree well with the ones obtained 
with the Gogny-D1S EDF \cite{rayner10odd-1,CEA}. For $N \approx$ 60 
isotones, the constrained Hartree-Fock plus BCS calculations, based on 
the Skyrme SLy4 \cite{sly} parameter set, predicted a trend similar to 
ours, i.e., an oblate-to-prolate shape transition between $N$=58 and 
60. A more gradual transition has been found within the relativistic 
mean-field framework based on the PC-PK1 parametrization and a density
dependent pairing \cite{mei2012}. For both Zr and Sr nuclei, an oblate (prolate) 
ground state minimum has been found for $N$=58 ($N$=60) in the 
framework of the Nilsson-Strutinsky method with a  deformed Woods-Saxon 
potential  and monopole pairing \cite{skalski1997}.


\subsection{Choice of the IBM configuration spaces}

\begin{figure*}[htb!]
\begin{center}
\includegraphics[width=0.7\linewidth]{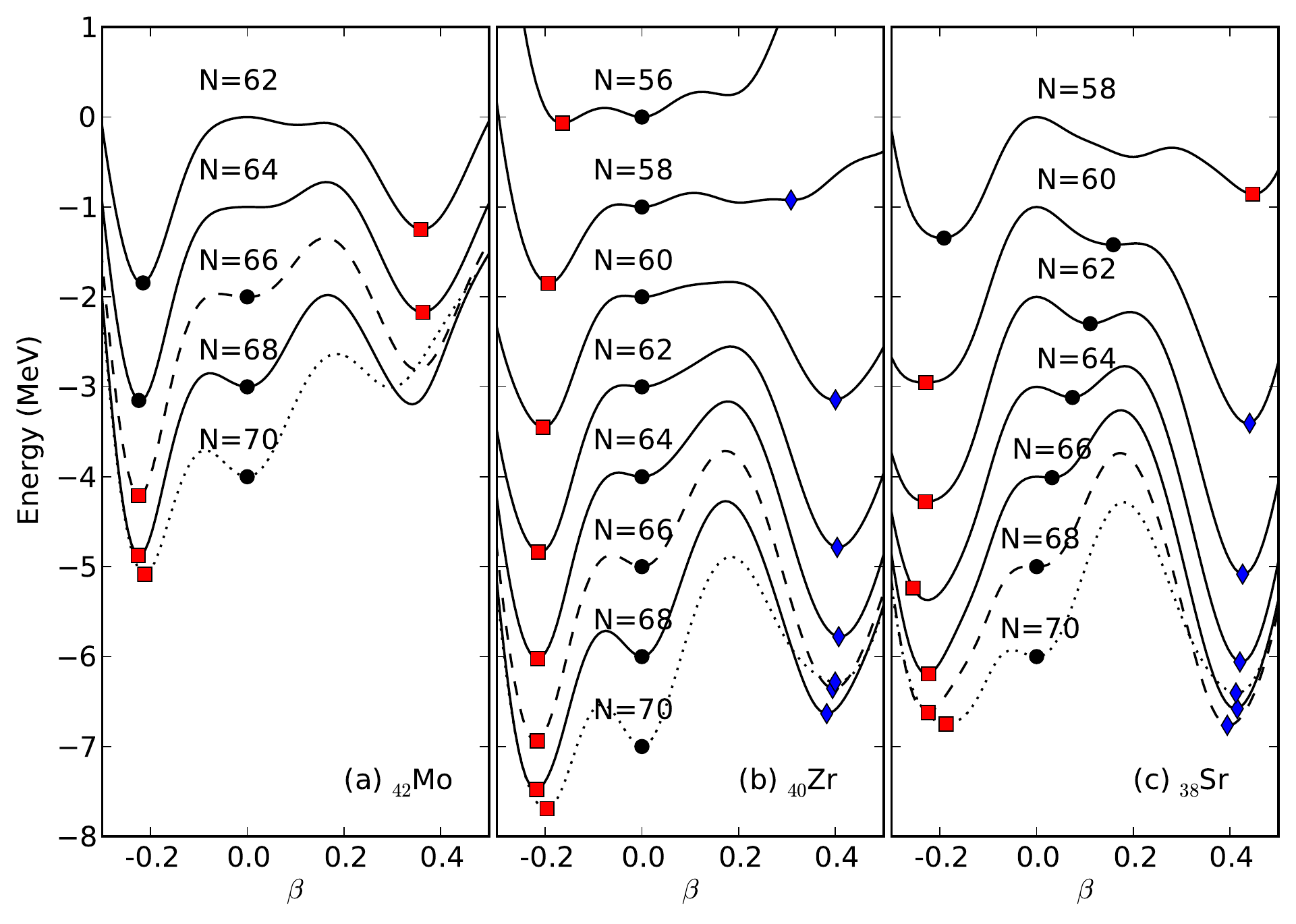}
\caption{(Color online) Projection of the HFB deformation energy
 surfaces along the $\beta$ axis for the $^{106-112}$Mo (a),
 $^{96-110}$Zr (b)
 and $^{96-108}$Sr (c) isotopes, where configuration mixing is taken into
 account in the corresponding IBM space. 
 Solid circles, squares and
 diamonds represent the $\beta$ values associated with the 
 $[n]$ ($n=0$, 1 and 2) unperturbed Hamiltonians,
 respectively.}
\label{fig:pes-axial}
\end{center}
\end{figure*}

We have performed the configuration mixing calculations in many of the
nuclei considered, where multiple mean-field minima have been observed in the
corresponding Gogny-D1M energy
surfaces. 
The criterion for choosing configuration spaces is whether a mean-field minimum
is clear (or deep) enough to constrain the corresponding unperturbed Hamiltonian. 
Using this criterion, configuration mixing calculations have been
carried out for heavier Mo nuclei and most of the Zr and Sr nuclei. 
However, such calculations have not been
carried out for all the Ru 
isotopes and the lightest nuclei in other isotopic chains, since the
microscopic Gogny-D1M energy surfaces only show one  
clear minimum (see Figs.~\ref{fig:pes-hfb-rumo} and \ref{fig:pes-hfb-zrsr}). 

For the sake of clarity, in 
what follows we summarize the configuration spaces for the nuclei
considered in this work.

\begin{itemize}
\item Ru: normal $[N_{\nu}\otimes (N_{\pi}=2)]$ configuration for all isotopes. 
\item Mo: normal $[N_{\nu}\otimes (N_{\pi}=1)]$ configuration for $^{96-102}$Mo,
      and $[N_{\nu}\otimes (N_{\pi}=1)]\oplus [N_{\nu}\otimes (N_{\pi}=3)]$ configuration
      for $^{104-112}$Mo. 
\item Zr: normal $[N_{\nu}\otimes (N_{\pi}=0)]$ configuration for $^{94}$Zr,
      $[N_{\nu}\otimes (N_{\pi}=0)]\oplus [N_{\nu}\otimes (N_{\pi}=2)]$ configuration
      for $^{96}$Zr, and $[N_{\nu}\otimes (N_{\pi}=0)]\oplus
      [N_{\nu}\otimes (N_{\pi}=2)]\oplus [N_{\nu}\otimes (N_{\pi}=4)]$ configuration
      for $^{98-110}$Zr. 
\item Sr: normal $[N_{\nu}\otimes (N_{\pi}=1)]$ configuration for $^{92-94}$Sr,
      $[N_{\nu}\otimes (N_{\pi}=1)]\oplus [N_{\nu}\otimes (N_{\pi}=3)]$ configuration
      for $^{96}$Sr, and $[N_{\nu}\otimes (N_{\pi}=1)]\oplus
      [N_{\nu}\otimes (N_{\pi}=3)]\oplus [N_{\nu}\otimes (N_{\pi}=5)]$ configuration
      for $^{98-108}$Sr. 
\end{itemize}

In Fig.~\ref{fig:pes-axial} the HFB energies of $^{104-112}$Mo, 
$^{96-110}$Zr and $^{96-108}$Sr are plotted as a function of the 
$\beta$ deformation parameter. Those are the nuclei where configuration 
mixing is taken into account in the corresponding IBM space. The 
$\beta$-coordinates  associated with the $[n]$ ($n=0$, 1 and 2) spaces 
are indicated in the plots. In panel (b) of the figure, we consider 
$^{100}$Zr as an illustrative example. In this case, the global minimum 
is oblate  with $\beta\approx -0.2$ and it is associated with the 
$[n=1]$ configuration. The second-lowest minimum on the prolate side is 
associated with the $[n=2]$ configuration and, finally, the third, 
almost  spherical, minimum is associated with the normal $[n=0]$ 
configuration. Within our framework, the ground state $0^+_1$ is mainly 
composed of the configuration associated with the global minimum, while 
the $0^+_2$ excited state is constructed mainly from the configuration 
associated with the second-lowest minimum. In the particular example of 
$^{100}$Zr, as shown later in Fig.~\ref{fig:frac}, the $0^+_1$ 
($0^+_2$) state is predominantly composed of the oblate (prolate) 
configuration.


\subsection{Mapped IBM deformation energy surface}

\begin{figure*}[htb!]
\begin{center}
\includegraphics[width=0.8\linewidth]{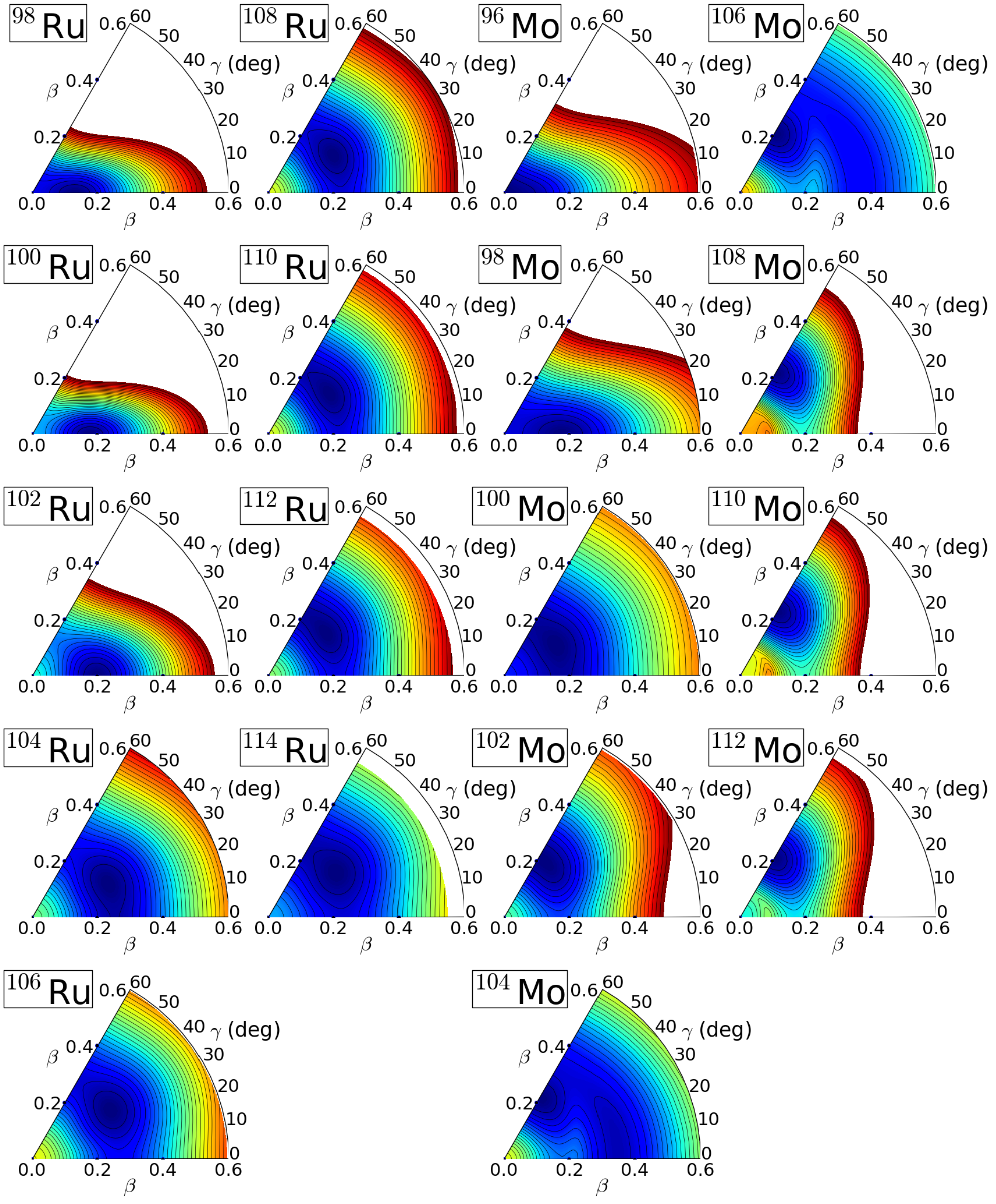}
\caption{(Color online) The same as in  Fig.~\ref{fig:pes-hfb-rumo}, but for the mapped
 IBM and for the Ru and Mo isotopes. }
\label{fig:pes-mapped-rumo}
\end{center}
\end{figure*}

\begin{figure*}[htb!]
\begin{center}
\includegraphics[width=0.8\linewidth]{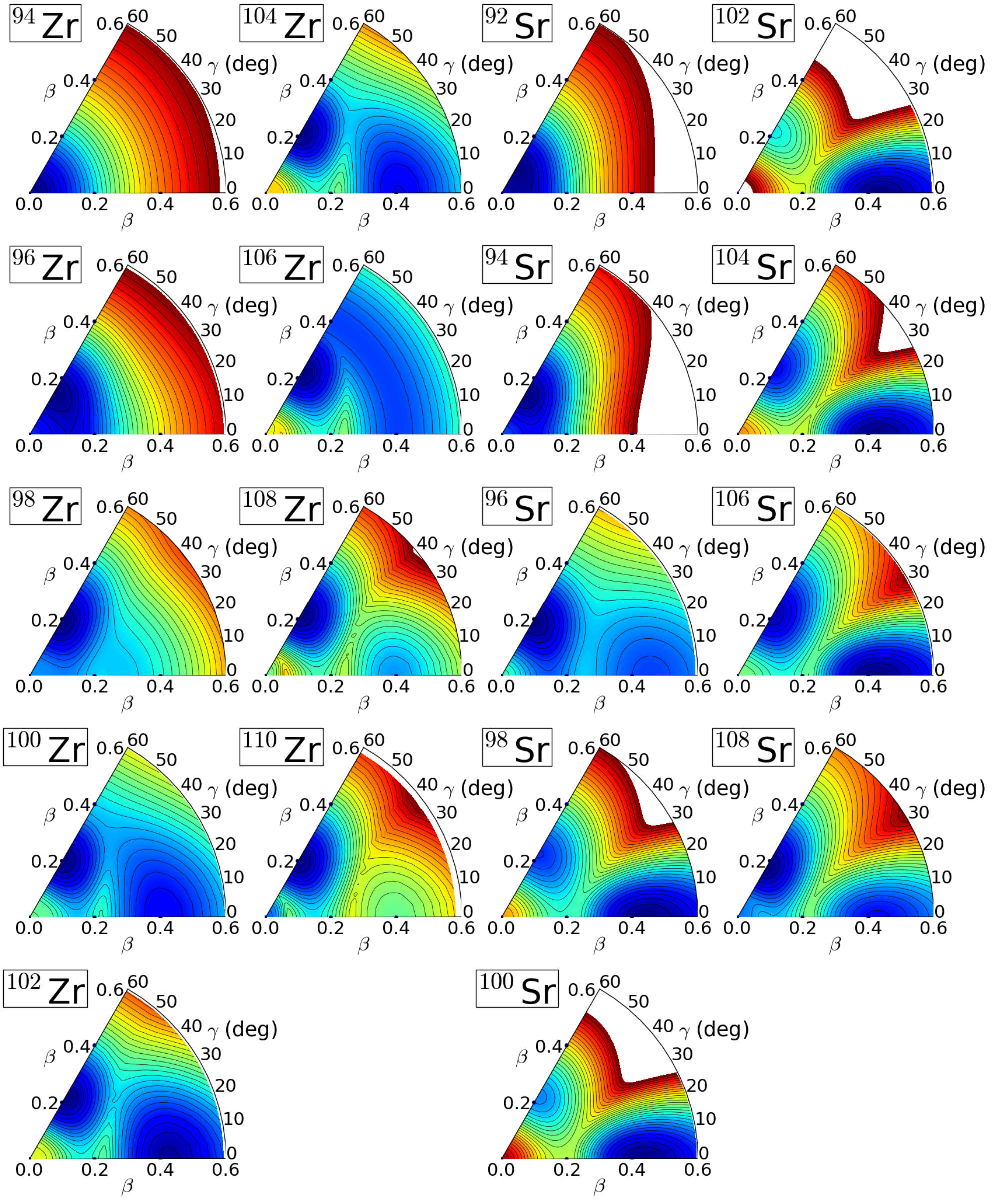}
\caption{(Color online) The same as in Fig.~\ref{fig:pes-hfb-rumo}, but for the mapped
 IBM and for the Zr and Sr isotopes. }
\label{fig:pes-mapped-zrsr}
\end{center}
\end{figure*}

In Figs.~\ref{fig:pes-mapped-rumo} and \ref{fig:pes-mapped-zrsr}, we 
have plotted the  IBM energy surfaces corresponding to the Hamiltonian 
Eq.~(\ref{eq:ham-cm}) whose parameters have been determined by  mapping 
the Gogny-D1M energy surfaces  shown in Figs.~\ref{fig:pes-hfb-rumo} 
and \ref{fig:pes-hfb-zrsr}.

Similar to the SCMF case, one observes for the Ru nuclei, shown in 
Fig.~\ref{fig:pes-mapped-rumo}, an evolution of the ground state 
deformation  from nearly spherical to triaxial  for 
$^{104}$Ru. The absolute minimum of the IBM surfaces becomes oblate for 
larger neutron numbers. The mapped surfaces obtained for Mo isotopes 
exhibit sharper minima than the ones found at the mean-field level. 
Moreover, for the neutron number $N\geq 66$ the mapped IBM surface is 
less $\gamma$-soft than the mean-field one. The IBM surfaces for Zr and 
Sr nuclei in Fig.~\ref{fig:pes-mapped-zrsr} also reproduce the overall 
HFB trend  as a function of the neutron number. In particular, they 
account for the onset of the strongly-deformed prolate shape around $N=60$ as 
well as a pronounced competition between the prolate and oblate minima 
for $60\le N \le 64$ in the case of Zr  and $58\le N \le 70$ for Sr nuclei.

As can be seen, the mapped energy surfaces 
reproduce the basic features of the mean-field ones. However, some 
discrepancies remain as the topology of the SCMF energy surfaces is richer than the IBM one. 
In particular, the mapped IBM surfaces tend to become flat in the 
region far from the minimum. There are essentially two reasons for this 
behavior: first, the analytical expressions of
Eqs.~(\ref{eq:pes-detail1}) and (\ref{eq:pes-detail2}) are  
too restricted to reproduce every detail of the original HFB energy 
surfaces and second, that the number and/or kind of bosons are rather 
limited within our IBM framework. 
One also observes substantial differences in the barriers between the
different minima. 
This is partly due to the adopted mixing
strength value $\omega=0.1$ MeV that may not be a proper choice for fully
reproducing the barriers. 
The employed value $\omega=0.1$ MeV is a guess based on our experience from
previous calculations \cite{nom12sc,nom13hg}. However, as will be shown later, it leads to a
reasonable description of spectroscopic properties.  
In order to fully reproduce the barriers, a much larger $\omega$ value
would be necessary. 
However, the larger $\omega$ value implies the stronger level repulsion between
the states and the model description would become unrealistically worse. 
For this reason, and since we are rather interested in describing an overall systematic trend of
the spectroscopic properties, we have tried not to
reproduce full details of the barriers and used a realistic mixing
strength $\omega=0.1$ MeV. 

Once the parameters of the Hamiltonian in Eq.~(\ref{eq:ham-cm}) have 
been fixed by the mapping procedure described in Sec.~\ref{sec:theory}, spectroscopic calculations are 
carried out to obtain excitation energies and transition probabilities.


\subsection{Parameters}

\begin{figure}[htb!]
\begin{center}
\includegraphics[width=\linewidth]{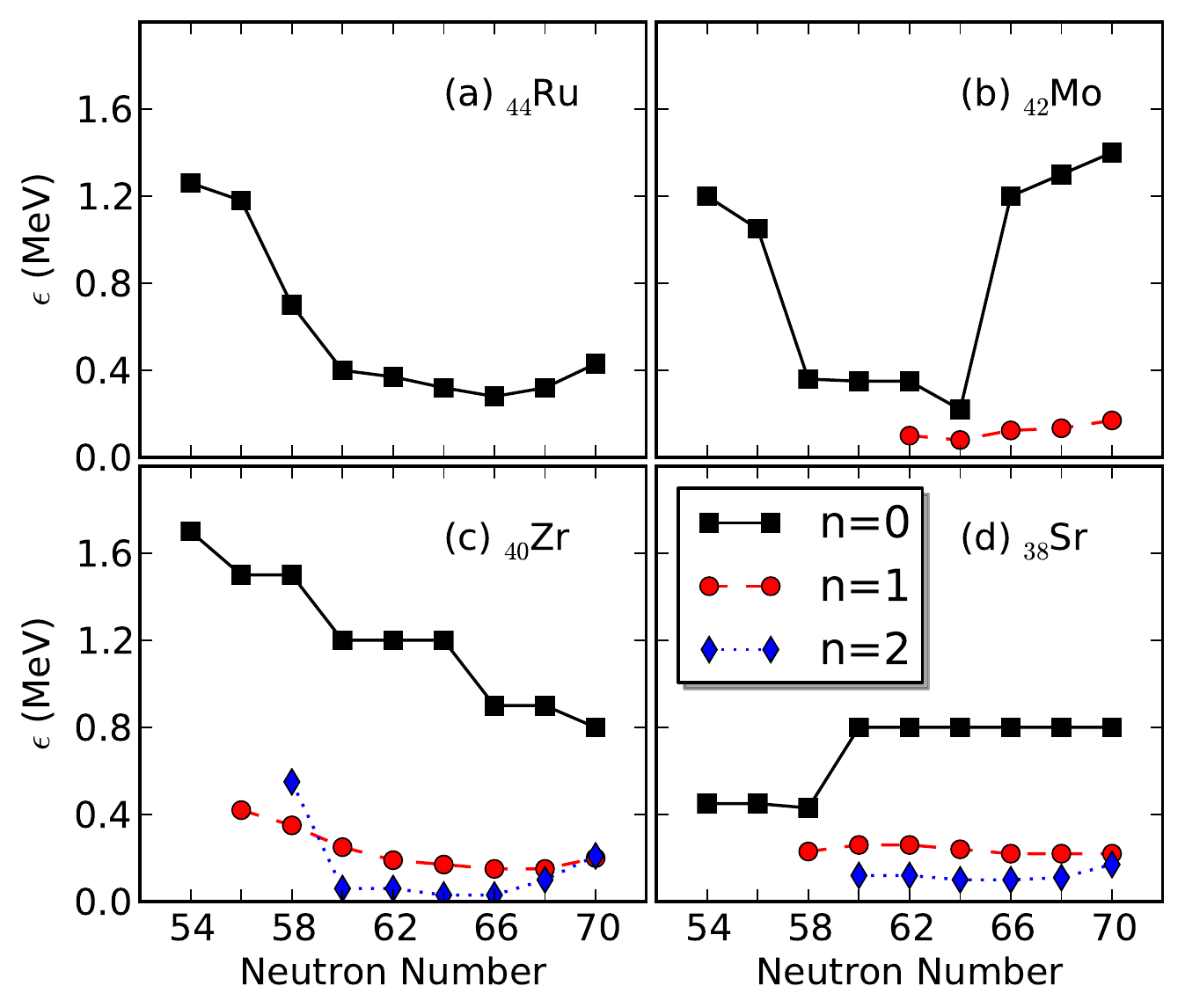}
\caption{(Color online) The parameter $\epsilon$ (in MeV) for the unperturbed
configuration spaces $[n]$ ($n=0$, 1 and 2) is plotted as a function of the
neutron number for the considered Ru, Mo, Zr and Sr isotopes. }
\label{fig:epsilon}
\end{center}
\end{figure}

\begin{figure}[htb!]
\begin{center}
\includegraphics[width=\linewidth]{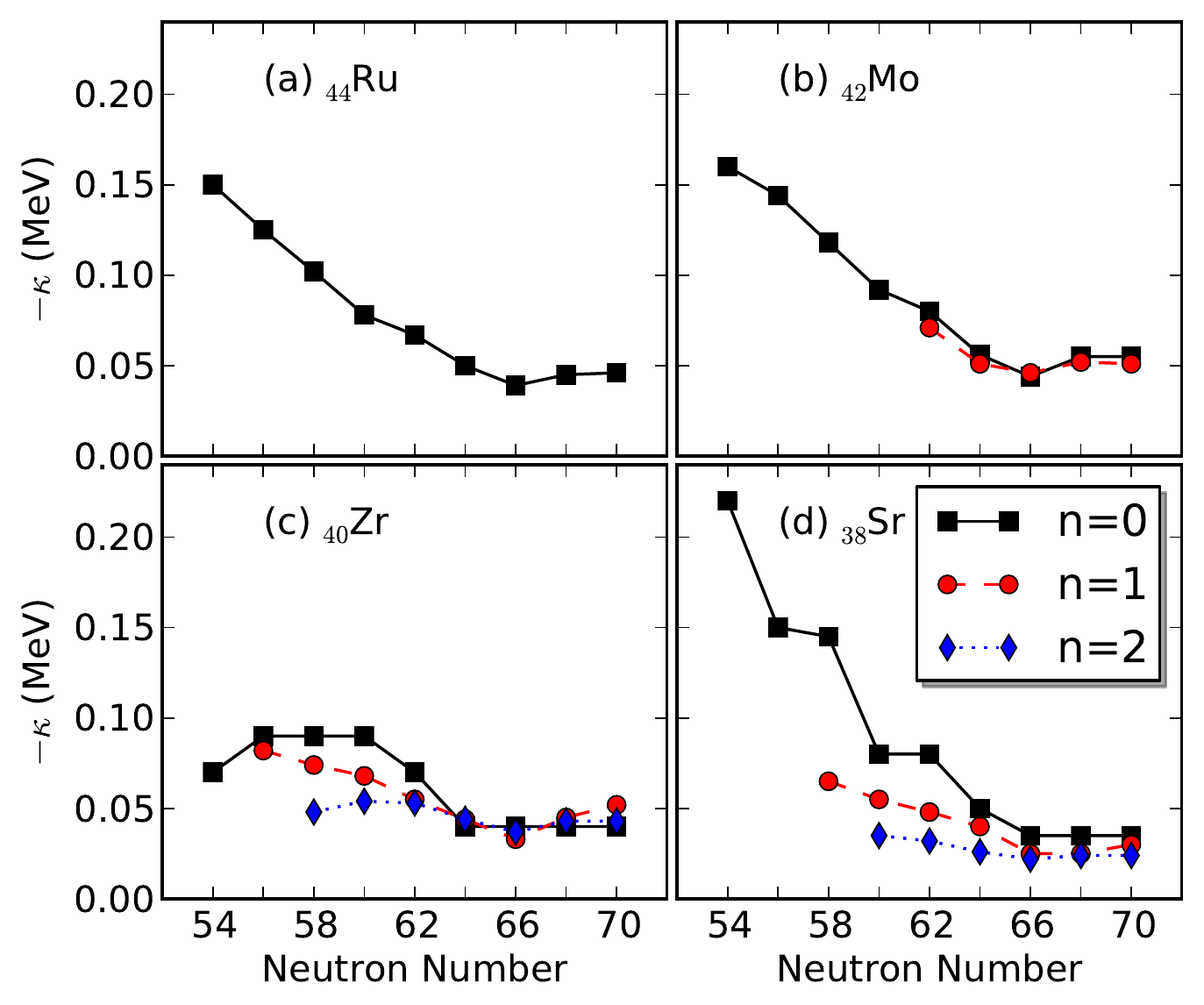}
\caption{(Color online) The same as in  Fig.~\ref{fig:epsilon},
but for the parameter $\kappa$ (in MeV). }
\label{fig:kappa}
\end{center}
\end{figure}

\begin{figure}[htb!]
\begin{center}
\includegraphics[width=\linewidth]{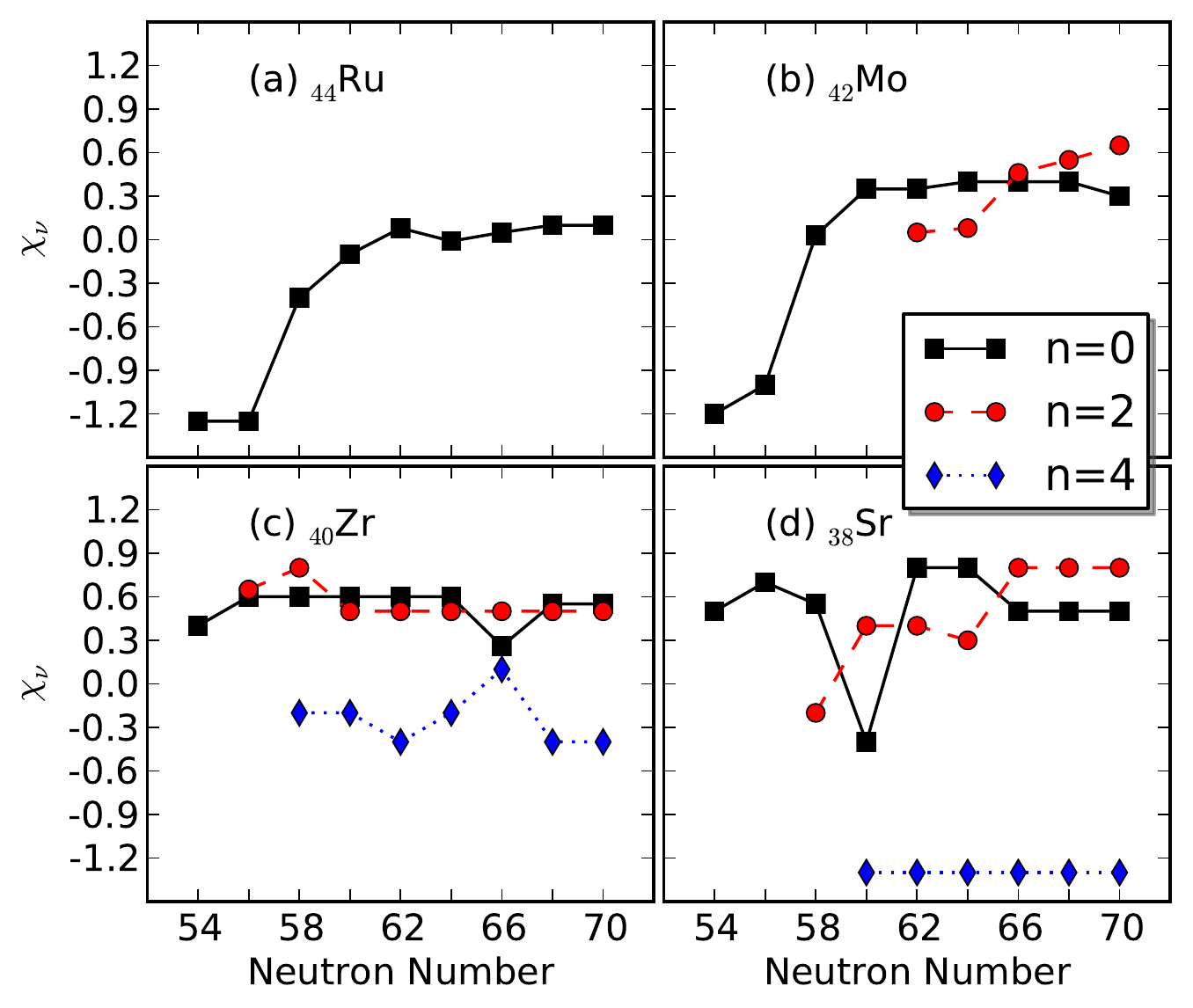}
\caption{(Color online) The same as in  Fig.~\ref{fig:epsilon},
but for the parameter $\chi_{\nu}$ (dimensionless). }
\label{fig:chn}
\end{center}
\end{figure}

\begin{figure}[htb!]
\begin{center}
\includegraphics[width=\linewidth]{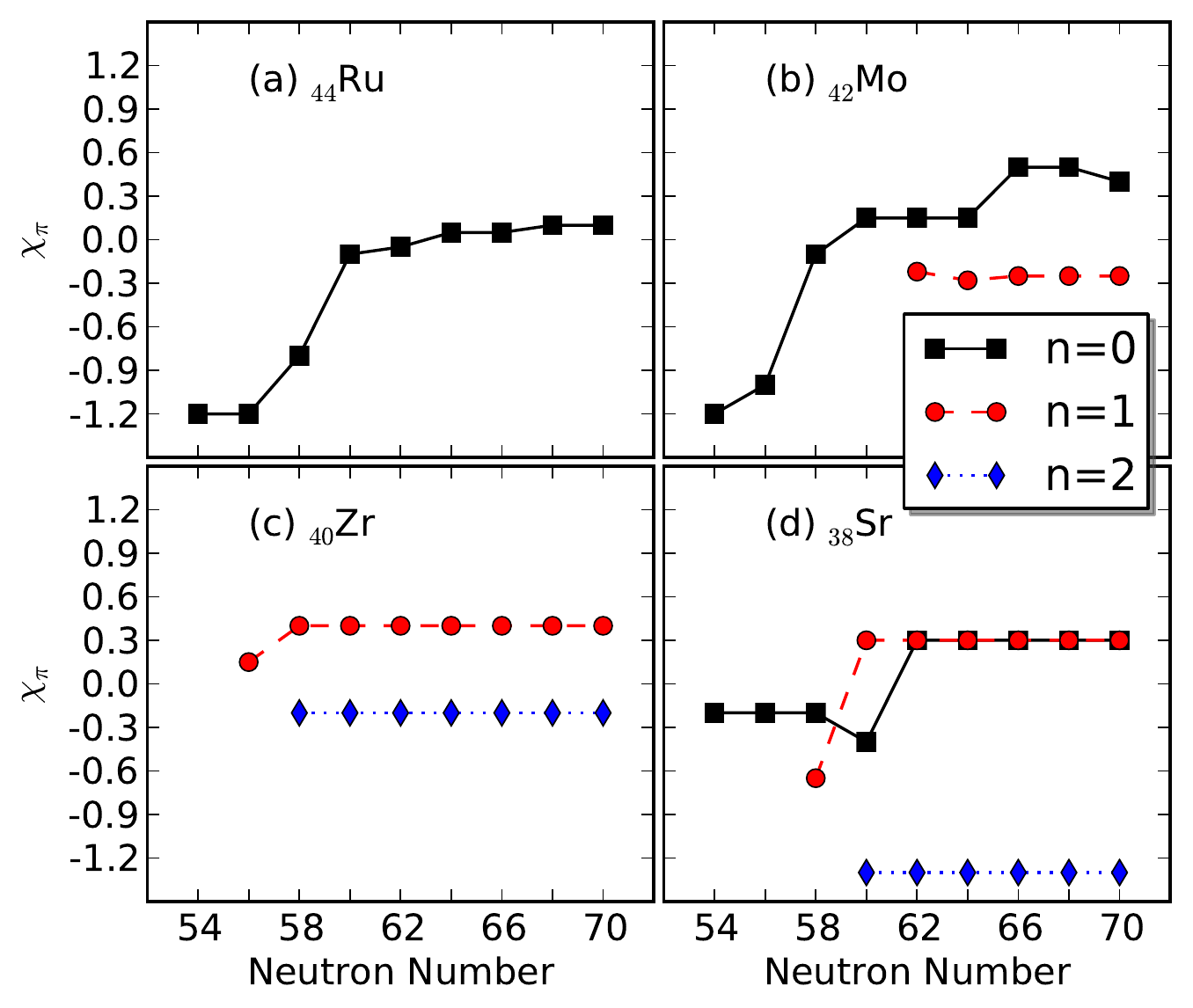}
\caption{(Color online) The same as in  Fig.~\ref{fig:epsilon},
but for the parameter $\chi_{\pi}$ (dimensionless). }
\label{fig:chp}
\end{center}
\end{figure}

\begin{figure}[htb!]
\begin{center}
\includegraphics[width=0.8\linewidth]{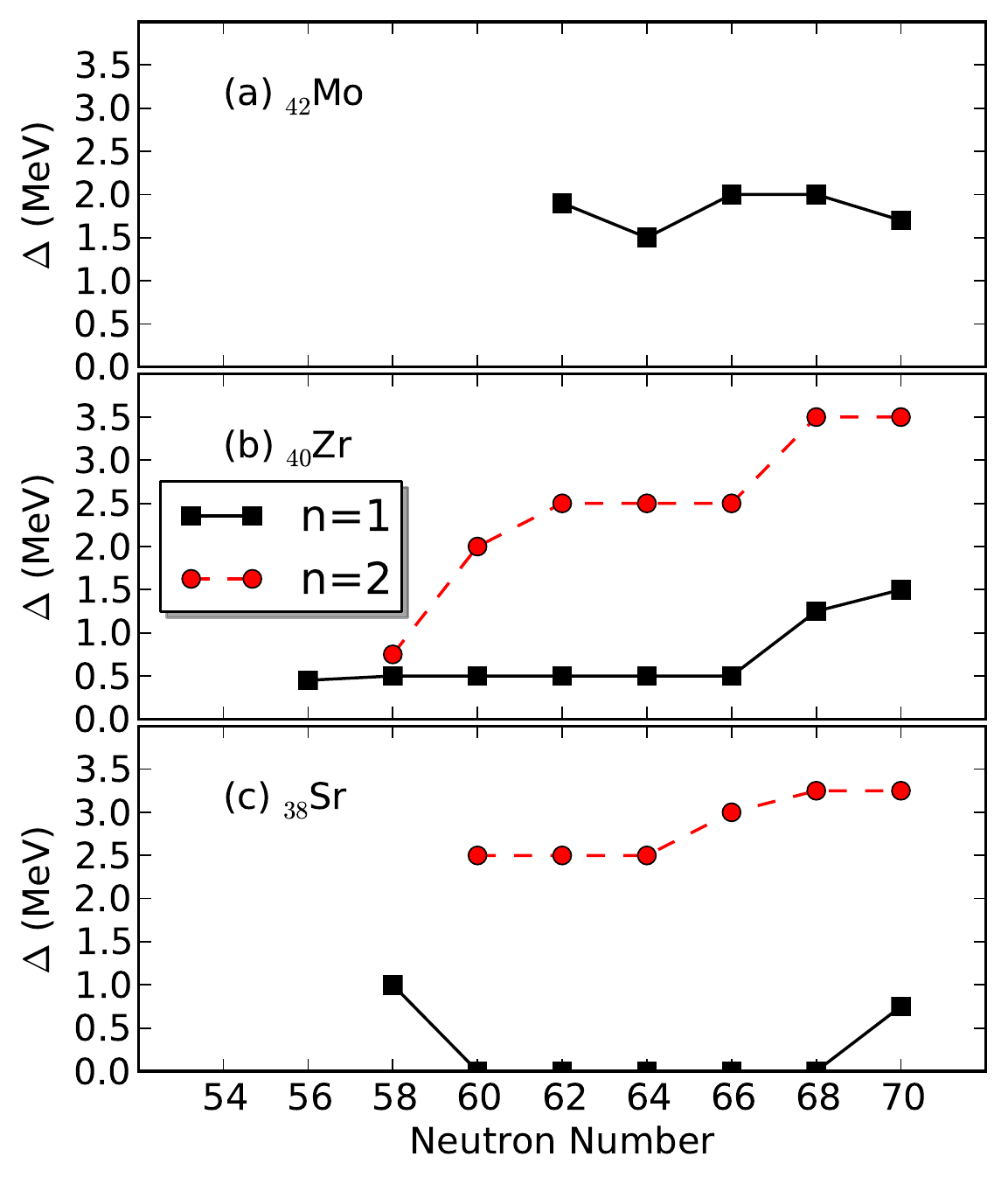}
\caption{(Color online) The energy offsets $\Delta_{1}$ and
$\Delta_{2}$ (in MeV) are plotted as a function of the neutron number for 
the Mo, Zr and Sr isotopes. }
\label{fig:delta}
\end{center}
\end{figure}

The parameters employed in our calculations for Ru, Mo, Zr and Sr 
isotopes are displayed in Figs.~\ref{fig:epsilon}-\ref{fig:delta}. They 
are obtained as a result of the mapping  procedure and their 
variations, as functions of the neutron number, reflect structural 
changes along the considered isotopic chains. We have not plotted the 
values of the strength $\kappa^{\prime}$ associated with the 
three-boson term as in most of the cases it is the same as the strength 
$\kappa$ for the quadrupole-quadrupole interaction. The exceptions are 
$^{104-114}$Ru and $^{100-102}$Mo. Their $\kappa^{\prime}$ values are 
0.25, 0.25, 0.12, 0.08, 0.10 and 0.18 MeV for $^{104-114}$Ru and the 
constant value  0.50 MeV for $^{100-102}$Mo. 

In panels (a) and (b)  of Fig.~\ref{fig:epsilon}, we have plotted the 
parameter $\epsilon$ for Ru and Mo nuclei, respectively. For most of 
the nuclei only a single configuration has been used in the 
calculations. As can be seen, $\epsilon$ decreases gradually towards 
the middle of the major shell. 
For the Mo isotopes, the value of $\epsilon$ for the normal
configuration exhibits a jump between $N=64$ and $N=66$. 
This is so because the structure of the 
unperturbed Hamiltonian for the normal configuration changes from 
$N=64$ to 66: the normal configuration is associated with the oblate 
minimum for $N\leq 64$, whereas it is associated with the nearly 
spherical minimum for $N\geq 66$ (see, Fig.~\ref{fig:pes-hfb-rumo}). 

For the Zr nuclei, shown in panel (c) of the same figure, $\epsilon$ 
decreases in heavier isotopes. However, the $\epsilon$ values for the 
$[n=1]$ and $[n=2]$ configurations change little for $N\geq 60$. For 
the Sr isotopes, shown in panel (d), exception made of the jump from 
$N=58$ to 60, an almost constant value is obtained. The parameter  
$\kappa$, plotted in Fig.~\ref{fig:kappa}(a-d), exhibits a gradual 
decrease towards the mid-shell for all configurations which is 
consistent, with the general mass-number dependence of this parameter 
\cite{IBM,OAI,mizusaki1996}.

The parameters $\chi_{\nu}$ and $\chi_{\pi}$ are depicted in 
Figs.~\ref{fig:chn} and \ref{fig:chp}. A certain combination of 
those parameters reflects whether a nucleus is either prolate or oblate 
deformed. From Eqs.~(\ref{eq:ham-sg}), (\ref{eq:pes-detail1}) and (\ref{eq:pes-detail1-1}),
one sees that the $\gamma$ dependence of the 
unperturbed Hamiltonian is associated with the two terms proportional to 
$-\kappa(N_{\nu}\chi_{\nu}+N_{\pi}\chi_{\pi})\cos{3\gamma}$ in the 
quadrupole-quadrupole interaction $\hat Q\cdot\hat Q$ and 
$-\kappa^{\prime}\sin^2{3\gamma}$ in the three-boson term.  Since the 
$\hat Q\cdot\hat Q$ is attractive ($\kappa<0$) then the minimum turns 
out to be prolate (oblate) if $N_{\nu}\chi_{\nu}+N_{\pi}\chi_{\pi}<0$ 
($N_{\nu}\chi_{\nu}+N_{\pi}\chi_{\pi}>0$). The three-boson term becomes 
important only when the nucleus is $\gamma$-soft, i.e., 
$N_{\nu}\chi_{\nu}+N_{\pi}\chi_{\pi}\approx 0$. The development of a 
triaxial minimum then depends on the strength of the three-boson 
interaction ($\kappa^{\prime}$). For many of the Ru and Mo nuclei, both 
$\chi_{\nu}$ and $\chi_{\pi}$ are close  to zero for $N\geq 60$ (see, 
panels (a) and (b) of Figs.~\ref{fig:chn} and \ref{fig:chp}) and the corresponding 
$\kappa^{\prime}$ values are large enough as to produce a triaxial 
minimum. The energy surfaces for many of the heavier Ru and Mo isotopes 
(see, Figs~\ref{fig:pes-hfb-rumo} and \ref{fig:pes-mapped-rumo}) 
display shallow triaxial minima. From panels (c) and (d) of Figs.~\ref{fig:chn} and 
\ref{fig:chp}, the $\chi_{\nu}$ and $\chi_{\pi}$ values for most 
of the Zr and Sr isotopes are chosen so as not to change too much with 
neutron number. Many of the deformation energy surfaces for the Sr 
isotopes exhibit pronounced prolate minimum around $\beta=0.4$, which 
are associated with the $[n=2]$ configuration. Consequently, their 
$\chi_\nu$ and $\chi_\pi$ values for the $[n=2]$ configuration are 
notably large $\approx -1.3$, being close to the SU(3) limit of the 
IBM.

The offset energies $\Delta_{1}$ and $\Delta_{2}$ used in our 
configuration mixing calculations are shown in Fig.~\ref{fig:delta}. 
They are of the order of a few MeV. This is consistent with the results 
of previous studies \cite{nom12sc,nom13hg} in different regions of the 
nuclear chart.
However, we have chosen $\Delta_2=0$ for the Sr isotopes with $60\le
N\le 68$ (see, panel (c)), due to the peculiar topology of the corresponding HFB
surfaces. 


\subsection{Evolution of low-lying levels}

\begin{figure*}[htb!]
\begin{center}
\includegraphics[width=0.7\linewidth]{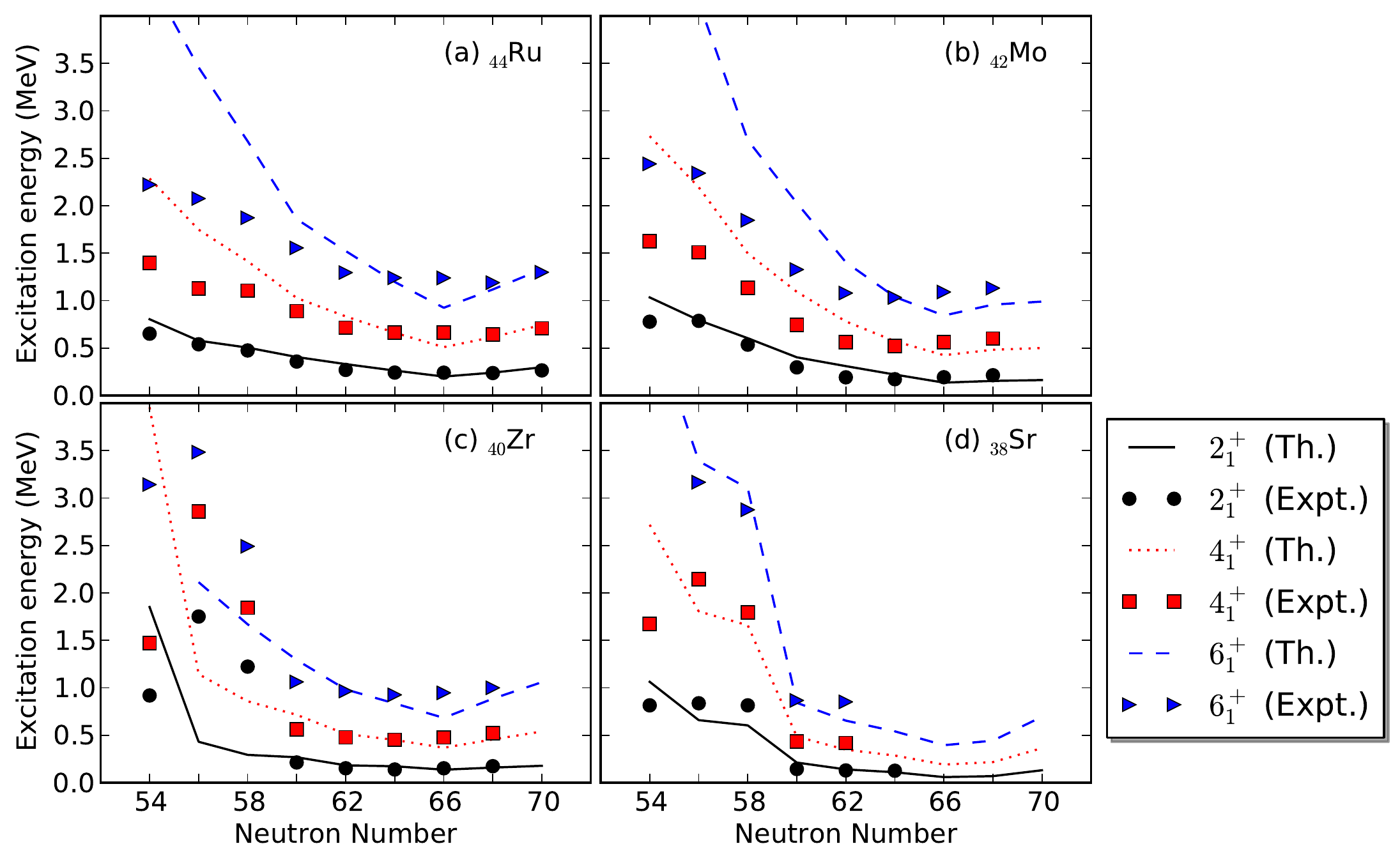}
\caption{(Color online) Evolution of the low-lying yrast states in the considered 
Ru, Mo, Zr and Sr isotopes as a function of the neutron number. Experimental 
data have been taken from Ref.~\cite{data}.}
\label{fig:gs}
\end{center}
\end{figure*}

\begin{figure*}[htb!]
\begin{center}
\includegraphics[width=0.7\linewidth]{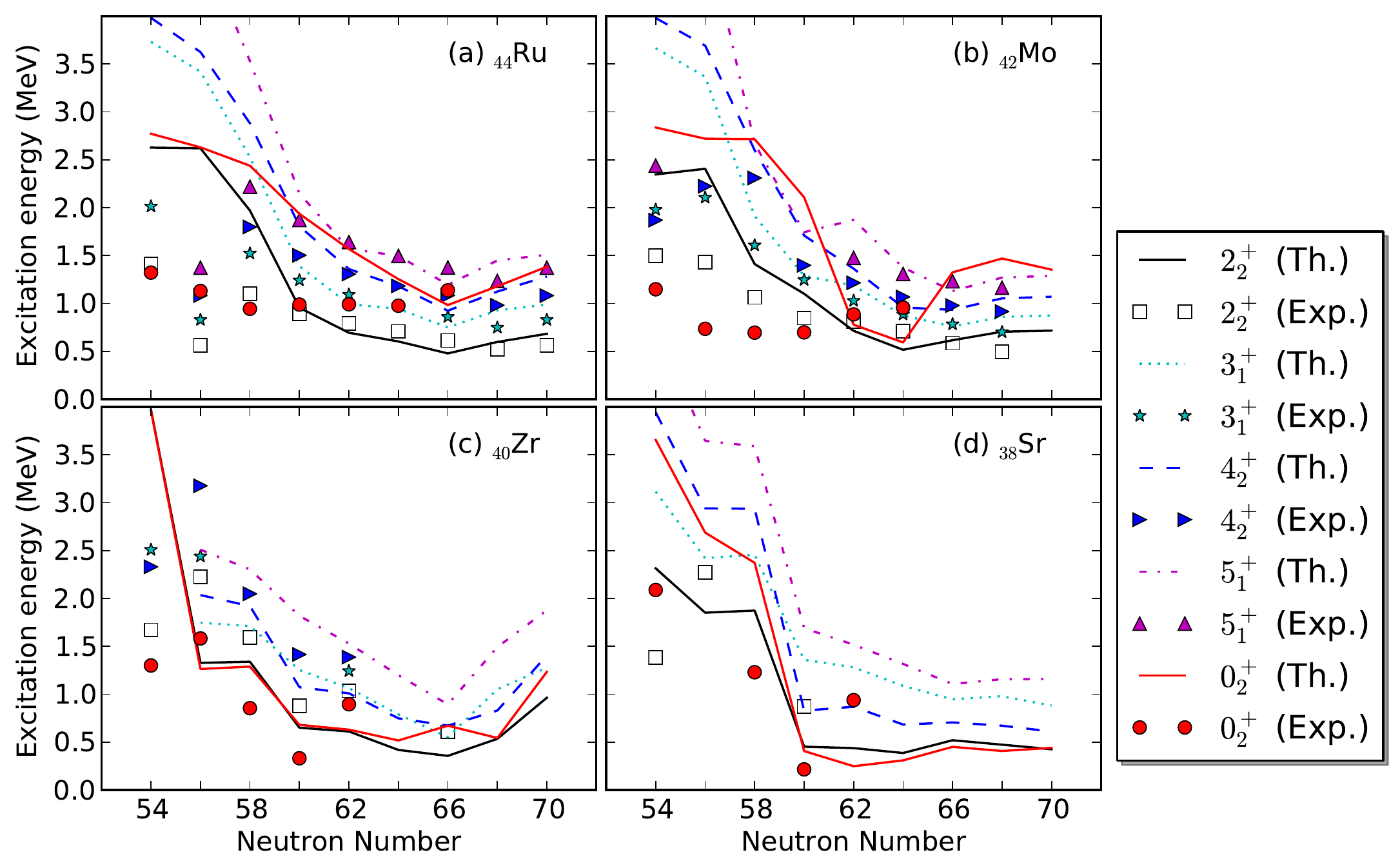}
\caption{(Color online) The same as in Fig.~\ref{fig:gs}, but
for the non-yrast states. }
\label{fig:qg}
\end{center}
\end{figure*}

The energies of some low-lying yrast and non-yrast states in 
$^{98-114}$Ru, $^{96-112}$Mo, $^{94-110}$Zr and $^{92-108}$Sr  are 
shown in Figs.~\ref{fig:gs} and \ref{fig:qg} as  functions of the
neutron number $N$. The energies of the yrast states decrease towards  
$N=66$ corresponding to mid-shell. They remain quite low for heavier Zr and Sr nuclei, 
reflecting a pronounced  collectivity. Experimentally, an abrupt change 
is observed in going from $N=58$ to 60 in the Zr and Sr chains. Although our 
calculations account reasonably well for the experimental trend in the 
low-energy yrast states, in particular for the  heavier ($N\geq 60$) 
isotopes, several deviations are also found: 

\begin{itemize}

 \item The energy levels near the neutron shell closure $N=50$
       are overestimated. This is due to the fact that the IBM model space,
       comprising only a finite number of $s$ and $d$ bosons, is not large
       enough to describe the energy levels near the closed shell. 

\item For Zr isotopes, the present calculation predicts that the yrast
      states change from
      $N=58$ to 60 gradually, whereas a much more drastic change is
      observed experimentally. This indicates that,
      in the present model calculation, these yrast states are rather similar
      in structure between the $^{98}$Zr and
      $^{100}$Zr nuclei. In both nuclei, the yrast states are mainly
      composed of the oblate $[n=1]$ configuration. Note, that the topology of
      the Gogny-D1M energy surface around this oblate minimum is rather 
      similar for the two nuclei. 

\item In the case of the Zr isotopes, the experimental high-lying yrast levels at $N=56$
      suggest the presence of a sub-shell closure. In our calculations, however, such    
      a feature is not reproduced. This is not surprising, since the present 
      IBM model space does not consider $N=56$ to be a shell closure and, as a consequence,  the energy levels 
      change rather gradually from $N=54$ to 56, as already found in previous  
      IBM calculations \cite{boyukata2010}. 

\end{itemize}

From the systematics of the non-yrast states depicted in 
Fig.~\ref{fig:qg}, one sees a typical $\gamma$-band sequence 
$2^+$, $3^+$, $4^+$, $5^+,\ldots$. The deviation with respect to the 
experimental data near $N=50$ could also be due to the limited number 
and/or types of bosons taken into account in our calculations. For both 
the Ru and Mo nuclei, the excitation energy of the $0^+_2$ state is 
generally overestimated as the intruder configuration is not considered 
in many of these nuclei, since the Gogny-HFB energy surface does not 
exhibit coexisting minima (see Fig.~\ref{fig:pes-hfb-rumo}). 
The $0^+_2$ energy level for the $N=62$ and 64 Mo nuclei is mostly coming from
the $\gamma$-soft minimum on the prolate side close in energy to
the oblate global minimum (see Fig.~\ref{fig:pes-mapped-rumo}) and is, 
consequently, notably low compared with other Mo nuclei.

For the nucleus $^{98}$Mo, already described using configuration mixing 
IBM calculations in Refs \cite{sambataro82,thomas2013}, we have obtained a 
pronounced deviation of the $0^+_2$ excitation energy with respect to 
the experimental value. Note that, at variance with previous results 
\cite{thomas2013}, our calculations on the deformation energy surfaces 
with the Gogny-D1M EDF (see, Fig.~\ref{fig:pes-hfb-rumo}) predict a 
single minimum around $\beta\approx 0.15$ (see, 
Fig.~\ref{fig:pes-hfb-rumo}). The structure of the Zr and Sr isotopes 
is characterized by  very low-lying excited $0^+$ states. In our 
calculations, the excited $0^+$ states are mainly dominated by the 
intruder configurations and their energies are similar to the energy 
difference between different mean-field minima. Our results suggest 
that for the Zr and Sr isotopes the $0^+_2$ excitation energy decreases 
from $N=58$ to $N=60$. This reflects the fact that in the mean-field 
calculations the second prolate minimum appears from $N=60$. On the 
other hand, the experimental $0^+_2$ excitation energy increases from 
$N=60$ to 62 while in our calculations it remains low.
  

\subsection{Wave functions for the $0^+_1$ and $0^+_2$ states}

\begin{figure*}[htb!]
\begin{center}
\includegraphics[width=0.7\linewidth]{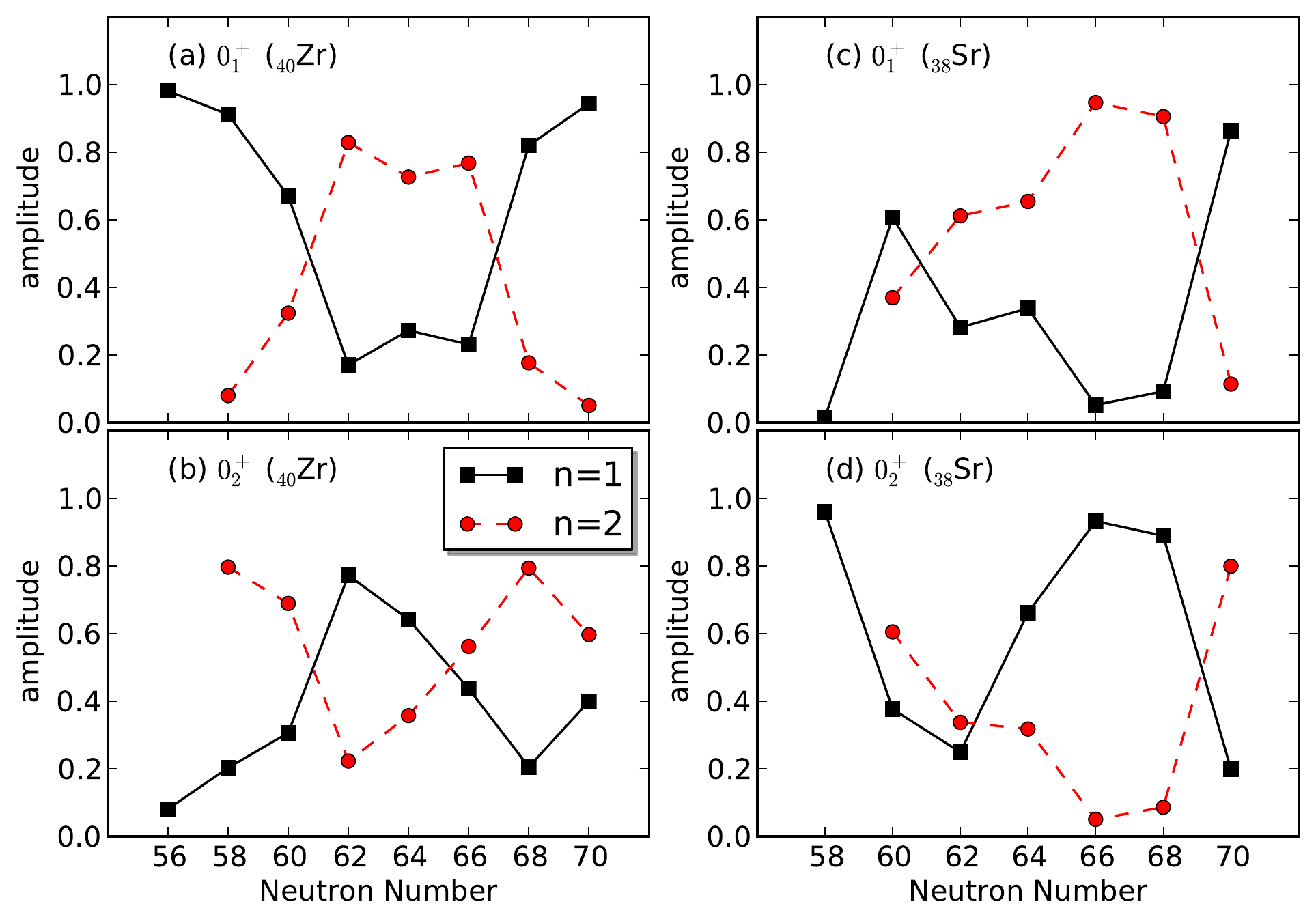}
\caption{(Color online) Amplitudes of the unperturbed $[n=1]$ and $[n=2]$
components in the wave functions of the $0^+_1$ and $0^+_2$ states of
$^{96-110}$Zr [Panels (a) and (b)] and $^{96-108}$Sr [Panels (c) and (d)] 
isotopes.}
\label{fig:frac}
\end{center}
\end{figure*}

To interpret the nature of the lowest two $0^+$ states, we have plotted 
in Fig.~\ref{fig:frac}, the amplitudes of the  unperturbed  $[n=1]$ and 
$[n=2]$ components in the wave functions of the $0^+_1$ and $0^+_2$ 
states for the  Zr and Sr nuclei. In most of the cases, the wave functions 
are composed predominantly of the oblate $[n=1]$ and prolate $[n=2]$ 
configurations, while the contribution of the normal $[n=0]$ 
configuration turns out to be  negligible. For this reason, we do not 
show the amplitude of the normal $[n=0]$ configuration. One should keep 
in mind that for $^{96}$Zr and $^{96}$Sr, the $[n=2]$ configuration is 
not included. 
So, the $0^+_1$ ($0^+_2$) state of  $^{96}$Zr ($^{96}$Sr) 
is almost purely made of the oblate (prolate) normal ($[n=1]$) 
configuration. 

From  Fig.~\ref{fig:frac}, one sees some characteristic features in the 
contents of the $0^+_1$ and $0^+_2$ wave functions  for $^{98-110}$Zr 
and $^{98-108}$Sr isotopes. From panel (a), one realizes that the 
$0^+_1$ state for $^{98,100}$Zr is  dominated by the oblate 
configuration. For $N=62-66$, on the contrary, the $0^+_1$ wave 
function is predominantly  prolate while for the $N=68$ and 70 
isotopes, the oblate configuration becomes dominant again. On the other 
hand, the systematics of the $0^+_2$ states, shown in panel (b) of the 
figure, reveals that the prolate configuration is dominant for  $N=58$ 
and 60, the oblate configuration makes major contribution for $N=62$ 
and 64 while the prolate configuration again dominates for $N=68$ and 
70. Similar results are found for the $0^+$ states in the Sr chain but 
more isotopes are found for which the prolate configuration becomes 
dominant in the ground state. The previous results are basically
consistent with the ones obtained at the HFB level (see,
Figs.~\ref{fig:pes-hfb-zrsr} and \ref{fig:pes-mapped-zrsr}), i.e., an
oblate ground state is observed for 
the isotopes with neutron numbers below $N\approx 58$ and beyond
$N\approx 68$, and a pronounced competition between an oblate and
a prolate minima in between.


\subsection{B(E2) systematics}

\begin{figure}[htb!]
\begin{center}
\includegraphics[width=\linewidth]{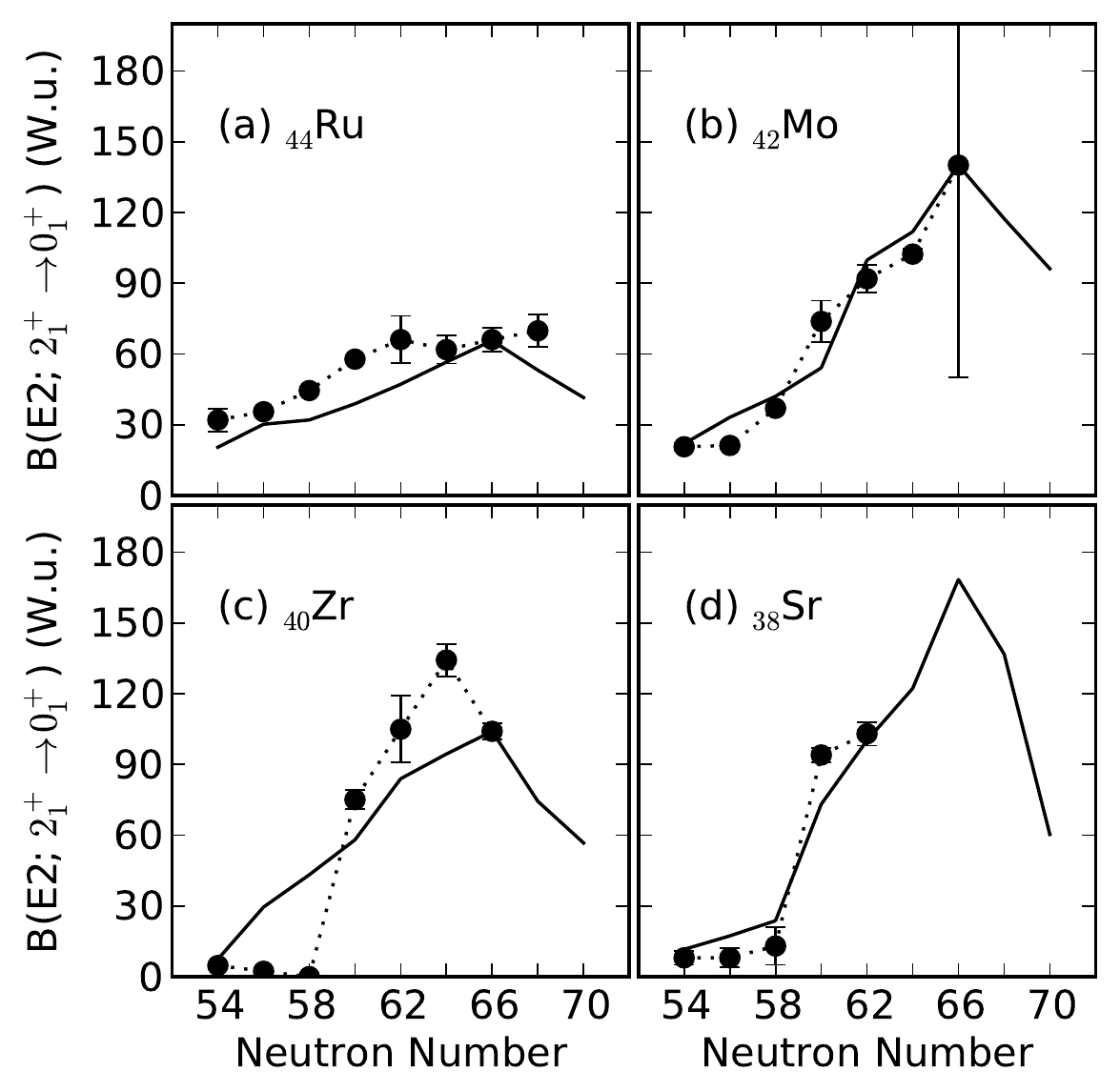}
\caption{$B$(E2; $2^+_1\rightarrow 0^+_1$)  transition strength 
(in Weisskopf units) for the considered Ru, Mo, Zr and Sr isotopes as a
function of the neutron number. Data have been taken from 
Refs.~\cite{data,browne2015}.}
\label{fig:2101}
\end{center}
\end{figure}

\begin{figure}[htb!]
\begin{center}
\includegraphics[width=\linewidth]{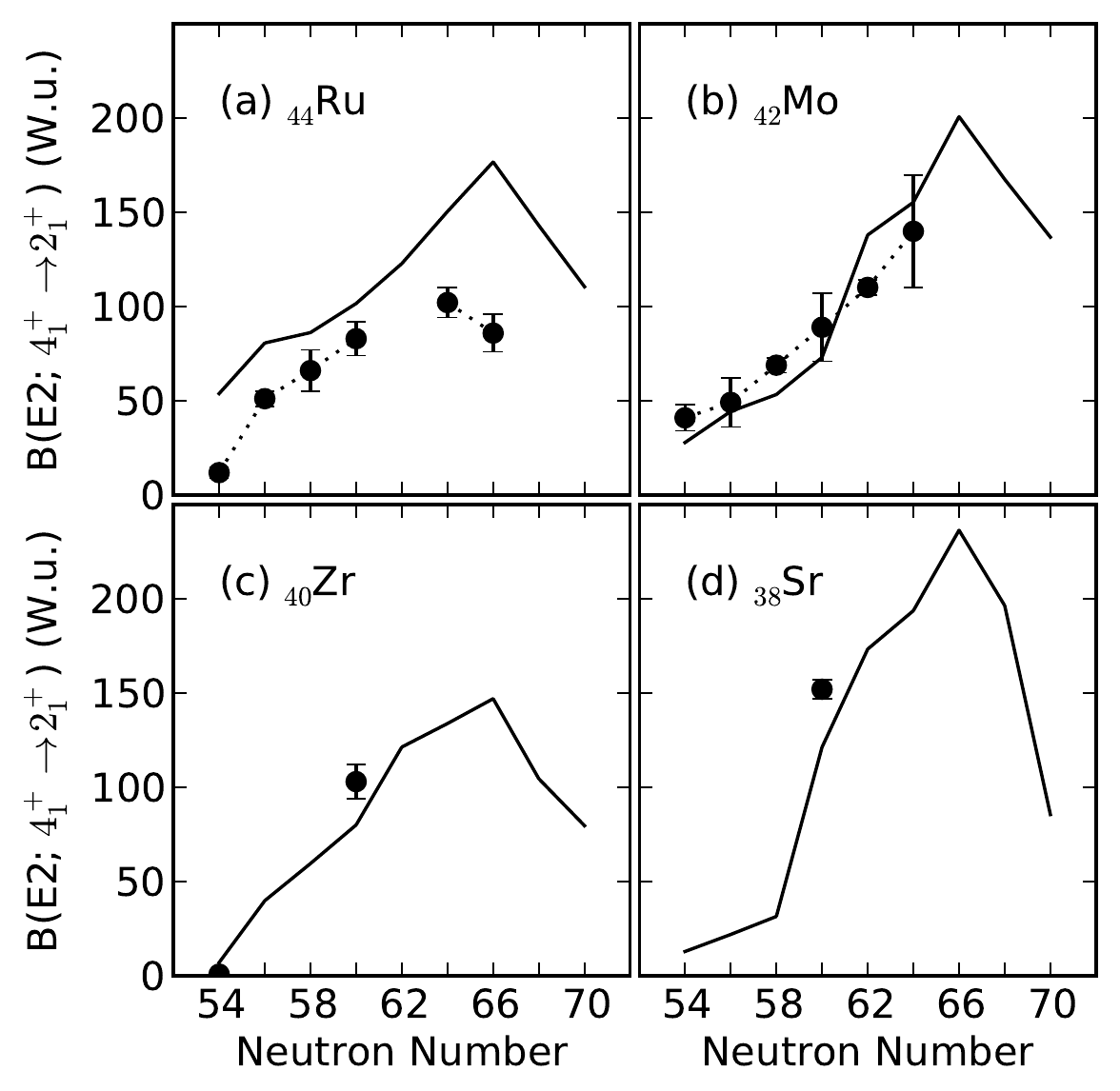}
\caption{The same as in  Fig.~\ref{fig:2101}, but for the $B$(E2; $4^+_1\rightarrow
2^+_1$)  transition strength. Data have been taken from Ref.~\cite{data}.}
\label{fig:4121}
\end{center}
\end{figure}

\begin{figure}[htb!]
\begin{center}
\includegraphics[width=\linewidth]{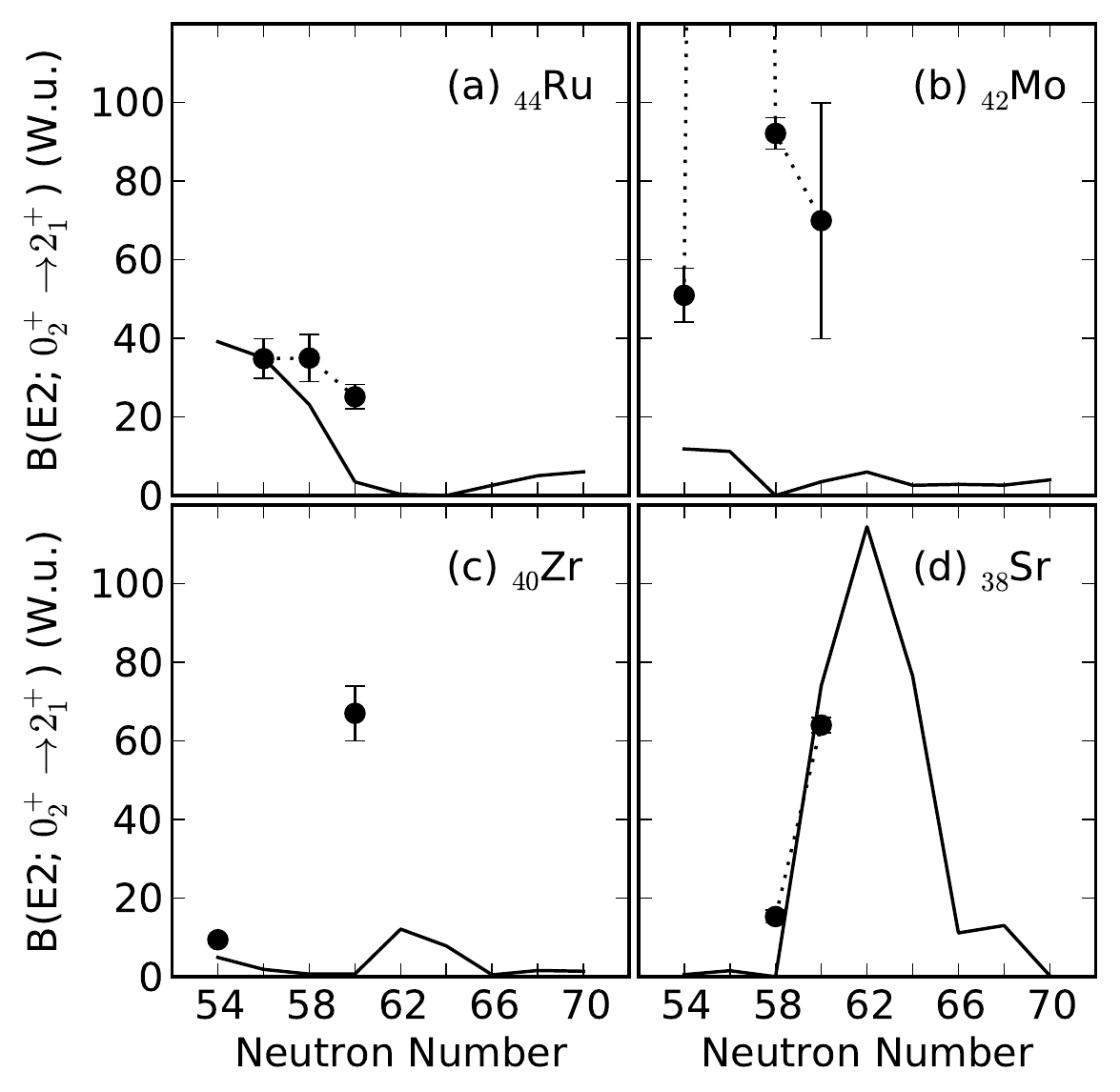}
\caption{The same as in  Fig.~\ref{fig:2101}, but for the $B$(E2; $0^+_2\rightarrow
2^+_1$)  transition strength. Data have been taken from Ref.~\cite{data}.}
\label{fig:0221}
\end{center}
\end{figure}

Figures \ref{fig:2101}, \ref{fig:4121} and \ref{fig:0221} show the
$B$(E2) transition strengths between the low-spin states, 
$B$(E2; $2^+_1\rightarrow 0^+_1$), 
$B$(E2; $4^+_1\rightarrow2^+_1$) and 
$B$(E2; $0^+_2\rightarrow 2^+_1$), respectively. 
The in-band   
$B$(E2; $2^+_1\rightarrow 0^+_1$) and 
$B$(E2; $4^+_1\rightarrow 2^+_1$) transitions become maximal around 
the neutron number $N=66$ corresponding to mid-shell
where the largest quadrupole collectivity is expected. 
In the case of the 
Zr isotopes, the experimental systematics suggests that the 
$B$(E2; $2^+_1\rightarrow 0^+_1$)  transition strength  remains small 
for $N=56$ and 58 (Fig.~\ref{fig:2101}) while ours increases gradually 
due to the  fact that in the considered IBM model space the $N=56$ 
sub-shell closure is not taken into account.

The inter-band $B$(E2; $0^+_2\rightarrow 2^+_1$) transition is shown in 
Fig.~\ref{fig:0221}. Near the vibrational limit, the value is 
comparable in magnitude to the $B$(E2; $2^+_1\rightarrow 0^+_1$) one. 
However, it becomes small in the deformed limit where such a transition 
is not allowed. Within this context, a vibrational-like behavior is 
suggested for the lighter Ru isotopes for which the deformation of the  
minimum in the corresponding mean-field energy surfaces is small. For 
heavier isotopes, this transition becomes small as the deformation 
becomes stronger. 

For Mo isotopes, the quite large experimental $B$(E2; $0^+_2\rightarrow 
2^+_1$) value of 1400~$\pm 200$ W.u. \cite{thomas2013} (not shown in 
Fig.~\ref{fig:0221}(b)) suggests a large mixing between the different 
intrinsic structures while the theoretical value is much smaller. The 
reason for the discrepancy is that for the lighter Mo nuclei 
configuration mixing is not taken into account in our calculations. A 
similar observation applies to Zr and Sr isotopes. For the former, the 
coupling between the $0^+_2$ and $2^+_1$ states is probably not strong 
enough to reproduce the experimental data while for the latter the 
trend is reasonably well described. 


\subsection{Detailed comparison of low-energy spectra}

So far, we have discussed  some key observables as  
functions of the neutron number. In this section, by means of the 
comparison with the available experimental data, we demonstrate that 
the mapping procedure  is also able to describe the detailed band 
structure and decay patterns for the $N=60$ isotones  $^{104}$Ru, 
$^{100}$Zr and $^{98}$Sr. To this end, the energy levels have 
been classified into bands according to their dominant E2 decay 
patterns. The level scheme for  $^{102}$Mo is strikingly similar to the 
one for $^{104}$Ru and is not discussed in what follows. 


\subsubsection{$^{104}$Ru}

\begin{figure*}[htb!]
\begin{center}
\includegraphics[width=0.65\linewidth]{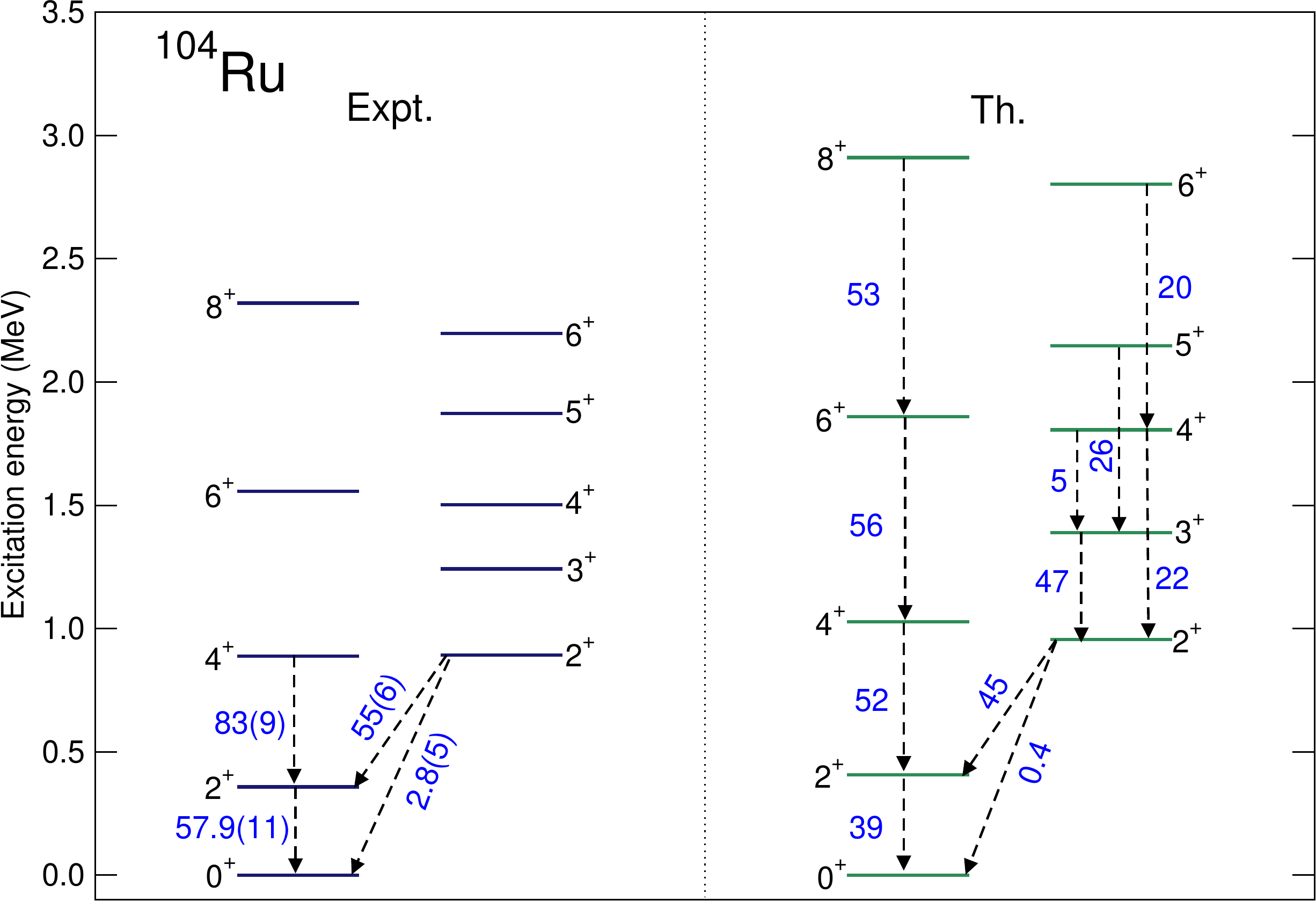}
\caption{(Color online) Low-energy level scheme for $^{104}$Ru. 
The numbers (in blue) near the arrows stand for the $B$(E2) values 
in Weisskopf units. Experimental data have been taken from Ref.~\cite{data}.}
\label{fig:104Ru}
\end{center}
\end{figure*}

The level scheme shown  in Fig.~\ref{fig:104Ru} for $^{104}$Ru 
corresponds to  a typical $\gamma$-soft spectra. The $2^{+}_2$, which 
is likely to be the band-head of the quasi-$\gamma$ band, lies close to 
the $4^{+}_1$ level. It also exhibits the E2 decay to the $2^+_1$  
state which is comparable to the $B(E2; 2^{+}_1\rightarrow 0^+_1)$ 
transition strength. The  energy spacing in the sequence $2^+_2$, 
$3^+_1$, $4^+_2$, $5^+_1$ and $6^+_2$ is rather constant. Like for 
other Ru isotopes, the quasi-$\gamma$ spectra obtained for $^{104}$Ru 
suggests that this system is somewhat in between the rigid-triaxial 
\cite{Davydov58} and the $\gamma$-unstable \cite{wilets1956} limits. 
Let us remark that the quasi-$\gamma$ band systematics can only be 
reproduced by including the three-boson term in the IBM Hamiltonian 
\cite{nom12tri}. The previous results compare well with the ones 
obtained using the 5D collective Hamiltonian approach, based on the 
deformed Nilsson potential and the Strutinsky's shell correction 
\cite{srebrny2006}. 


\subsubsection{$^{100}$Zr}

\begin{figure*}[htb!]
\begin{center}
\includegraphics[width=0.65\linewidth]{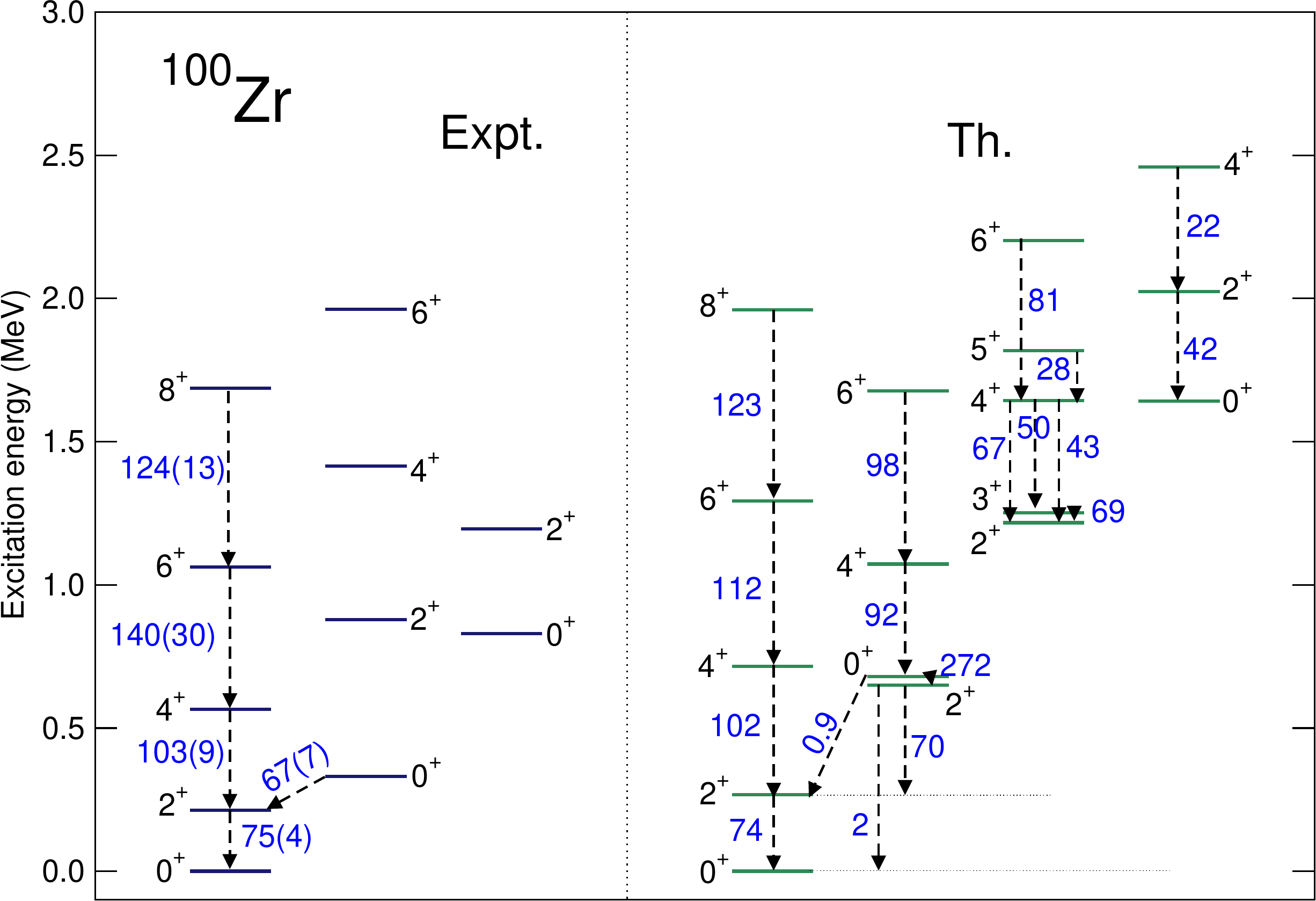}
\caption{(Color online) The same as in Fig.~\ref{fig:104Ru}, but
for the $^{100}$Zr nucleus. Data have been taken from Ref.~\cite{data}.}
\label{fig:100Zr}
\end{center}
\end{figure*}

Experimentally the nucleus $^{100}$Zr, shown in Fig.~\ref{fig:100Zr}, 
is characterized by the  $0^+_2$  and $2^+_1$ levels being rather close 
in energy and connected by a strong 
E2 transition probability of 67 $\pm 7$ W.u. On the other hand, our 
calculations overestimate the $0^+_2$ excitation energy, which is much 
higher than the $2^+_1$ energy level and also higher than the 
experimental counterpart and is even above the $2^+_2$ level. This is 
probably a consequence of strong level repulsion between the $0^+$ 
states, since the mixing strength $\omega=0.1$ MeV might be too large 
for this particular example. Also, the calculated $B$(E2; 
$0^+_2\rightarrow 2^+_1$) = 0.9 W.u. is too small compared with the 
experimental data. The too small E2 transition strength reflects that 
these states have different characters: 31 \% (69 \%) of the wave 
function of the $0^+_2$ ($2^+_1$) state is composed of the oblate 
$[n=1]$ configuration. 

Our calculations predict a third-lowest band based on the $2^+_3$ 
state, which resembles a quasi-$\gamma$ band  but shows the level 
structure of the rigid-triaxial rotor \cite{Davydov58}. The band built 
on the $0^+_3$ state is also overestimated, again due to strong level 
repulsion between the $0^+$ states.  


\subsubsection{$^{98}$Sr}

\begin{figure*}[htb!]
\begin{center}
\includegraphics[width=0.65\linewidth]{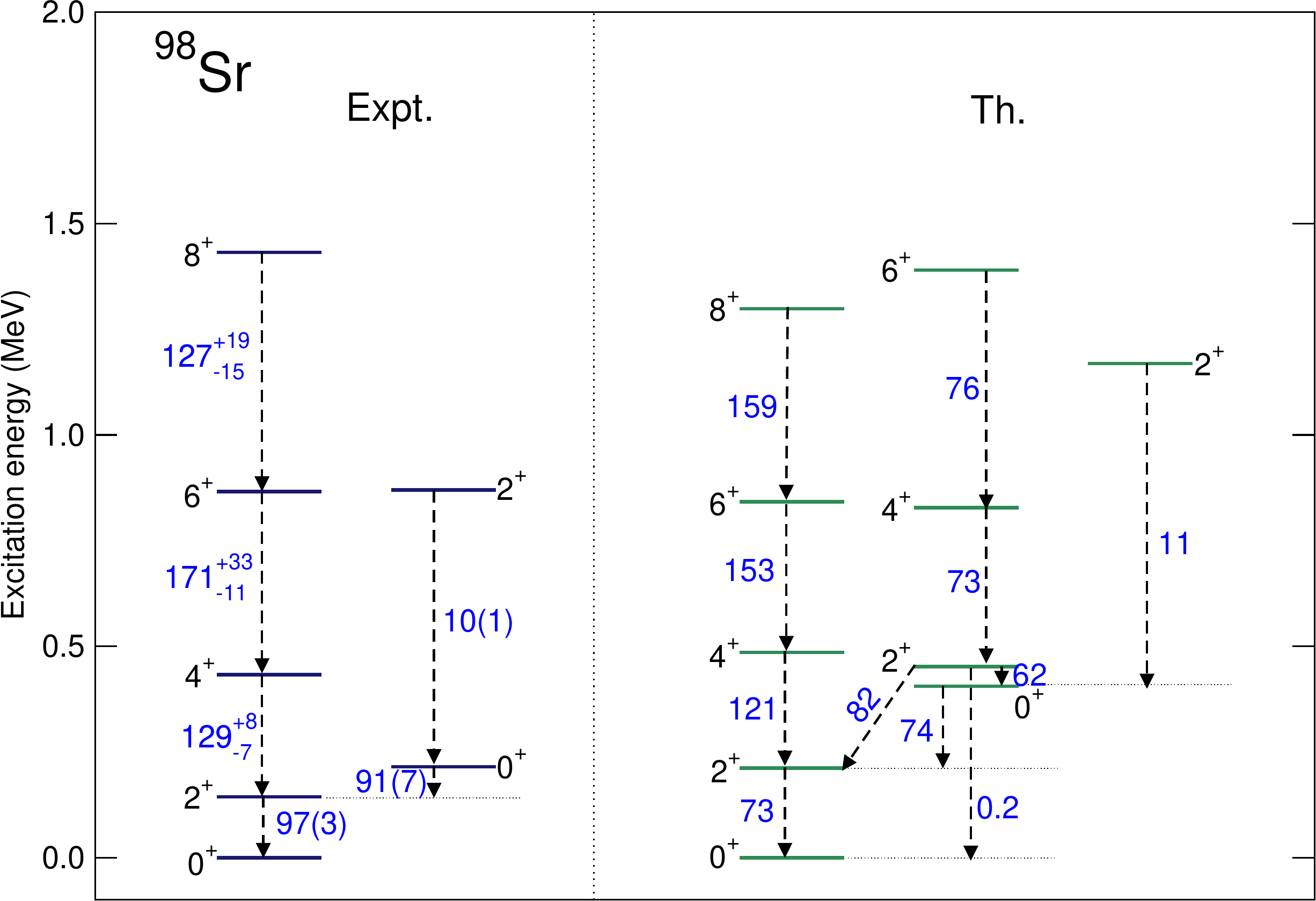}
\caption{(Color online) The same as in  Fig.~\ref{fig:104Ru}, but
 for the $^{98}$Sr nucleus. Data have been taken from Refs.~\cite{clement2016}.}
\label{fig:98Sr}
\end{center}
\end{figure*}

\begin{table}[hb!]
\caption{\label{tab:wf-98Sr} Fraction (in per cent units \%) of the three configurations 
$[n=0]$, $[n=1]$ and $[n=2]$ in the wave functions of 
the low-lying states of $^{98}$Sr  shown in Fig.~\ref{fig:98Sr}. }
\begin{center}
\begin{tabular*}{\columnwidth}{p{1.0cm}p{2.5cm}p{2.5cm}p{2.5cm}}
\hline\hline
\textrm{} &
\textrm{$[n=0]$}&
\textrm{$[n=1]$}&
\textrm{$[n=2]$}\\
\hline\hline
$0^+_1$ & 2.4 & 60.6 & 37.0 \\
$0^+_2$ & 1.9 & 37.6 & 60.5 \\
$2^+_1$ & 0.9 & 39.3 & 59.9 \\
$2^+_2$ & 1.0 & 67.2 & 31.9 \\
$2^+_3$ & 4.7 & 75.9 & 19.5 \\
$4^+_1$ & 0.2 & 16.8 & 83.0 \\
$4^+_2$ & 0.8 & 85.6 & 13.6 \\
$6^+_1$ & 0.0 & 8.2  & 91.8 \\
$6^+_2$ & 0.4 & 90.9 &  8.7 \\
$8^+_1$ & 0.0 & 4.6  & 95.4 \\
\hline
\end{tabular*}
\end{center}
\end{table}

Figure~\ref{fig:98Sr} depicts the level scheme for the nucleus  
$^{98}$Sr. The  fractions of the near-spherical $[n=0]$, oblate $[n=1]$ 
and prolate $[n=2]$ configurations introduced for this nucleus are 
given in Table~\ref{tab:wf-98Sr}. Though the level scheme is somewhat 
similar to the one obtained for $^{100}$Zr, the agreement with the 
experiment is better. From Table~\ref{tab:wf-98Sr}, one sees that the  
$0^+_1$ ($0^+_2$) state is dominated by the oblate $[n=1]$ (prolate 
$[n=2]$)  configuration. However, the prolate $[n=2]$ (oblate $[n=1]$) 
configuration becomes more dominant for higher angular momentum, which 
is compatible with the empirical systematics \cite{park2016}. The 
excitation energy of the $0^+_2$ state is overestimated but the strong 
decay to the $2^+_1$ state is consistent with the experiment and 
reflects the large mixing between these two states. As can be seen from 
Table~\ref{tab:wf-98Sr}, the content of the corresponding wave 
functions is similar. Experimentally there is a large energy gap 
between the $2^+_2$ and $0^+_2$ states. However, in our calculations 
both states are close in energy and connected by a strong E2 decay rate 
(62 W.u.). On the other hand, the theoretical $B$(E2; $2^+_3\rightarrow 0^+_2$) 
value of 11 W.u. agrees better with the experimental $B$(E2;
$2^+_2\rightarrow 0^+_2$) value. Let us mention that 
the quality of the agreement between our results and the experimental 
data is similar to the one obtained recently using the 5D collective 
Hamiltonian approach based on the Gogny-D1S EDF  \cite{clement2016}.


\subsection{Sensitivity tests \label{sec:test}}

\begin{figure}[htb!]
\begin{center}
\includegraphics[width=\linewidth]{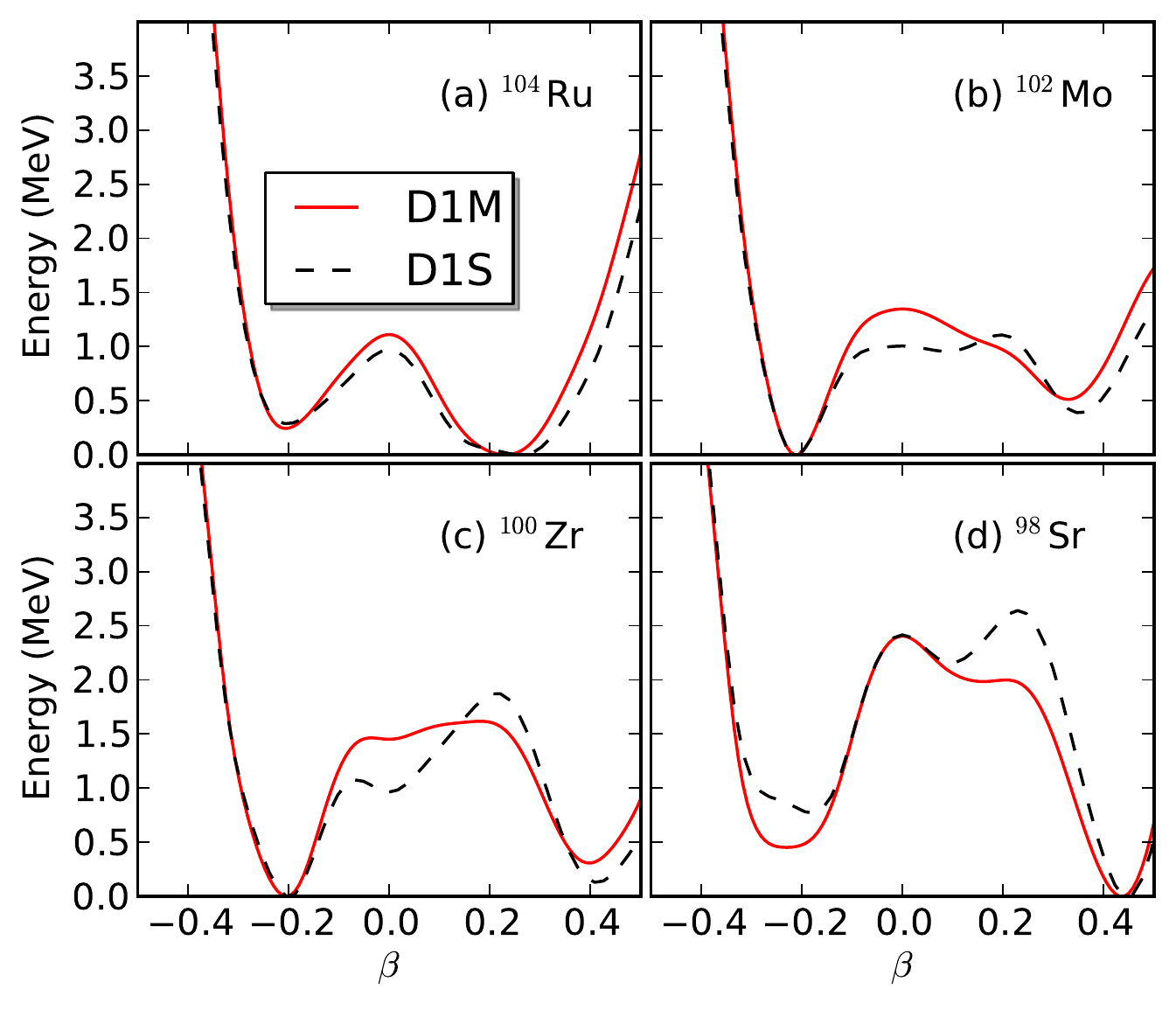}
\caption{(Color online) The SCMF deformation energy curves for the $N=60$
 isotones, (a) $^{104}$Ru, (b) $^{102}$Mo, (c) $^{100}$Zr and (d)
 $^{98}$Sr as functions of axial deformation parameter $\beta$ (with $\gamma=0^{\circ}$),
 calculated with the Gogny D1S and D1M parametrizations. }
\label{fig:d1sd1m_beta}
\end{center}
\end{figure}

\begin{figure}[htb!]
\begin{center}
\includegraphics[width=\linewidth]{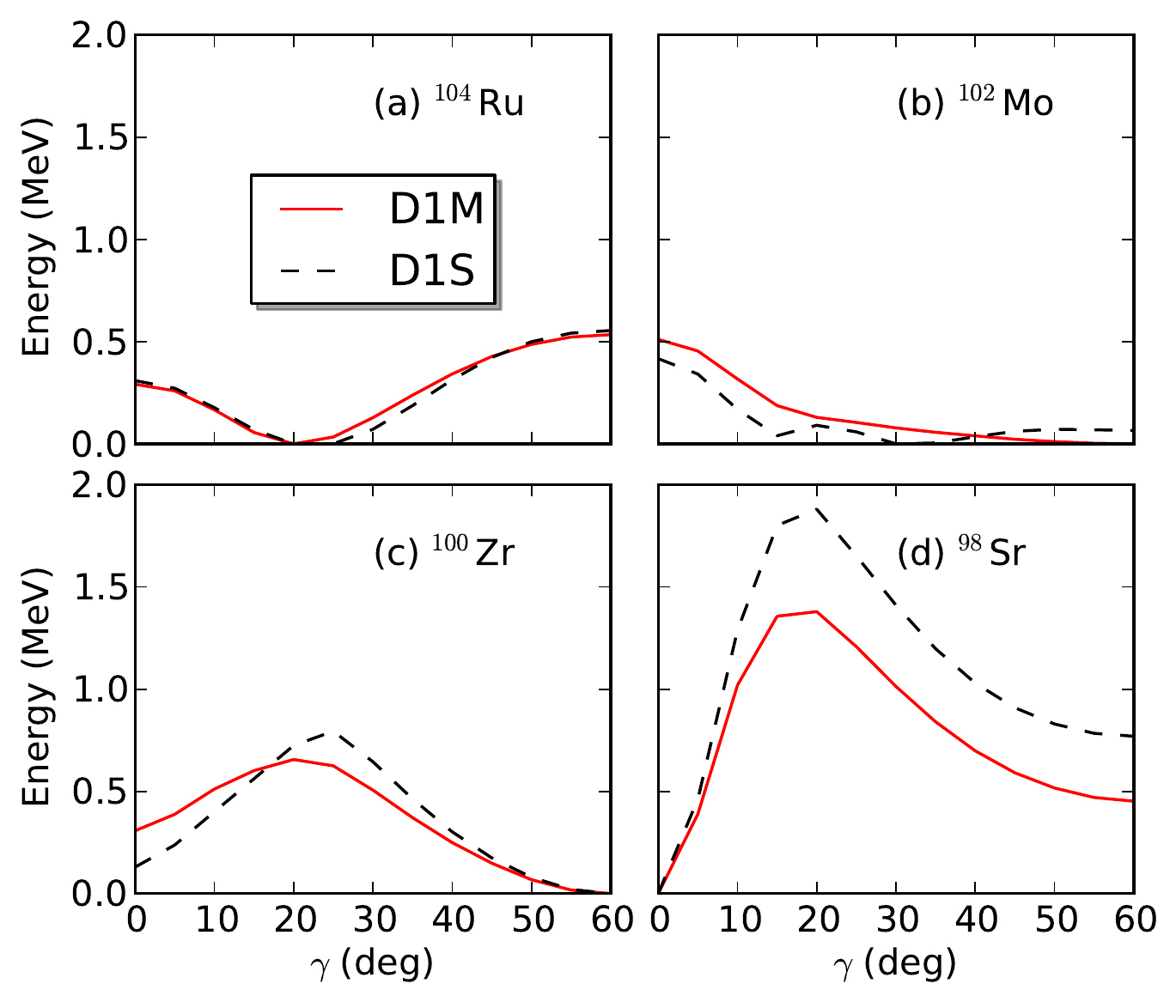}
\caption{(Color online) The same as in Fig.~\ref{fig:d1sd1m_beta}, but as
 functions of non-axial deformation parameter $\gamma$ with a $\beta$
 value corresponding to the minimum  at each $\gamma$ value. }
\label{fig:d1sd1m_gamma}
\end{center}
\end{figure}

\begin{figure*}[htb!]
\begin{center}
\includegraphics[width=0.65\linewidth]{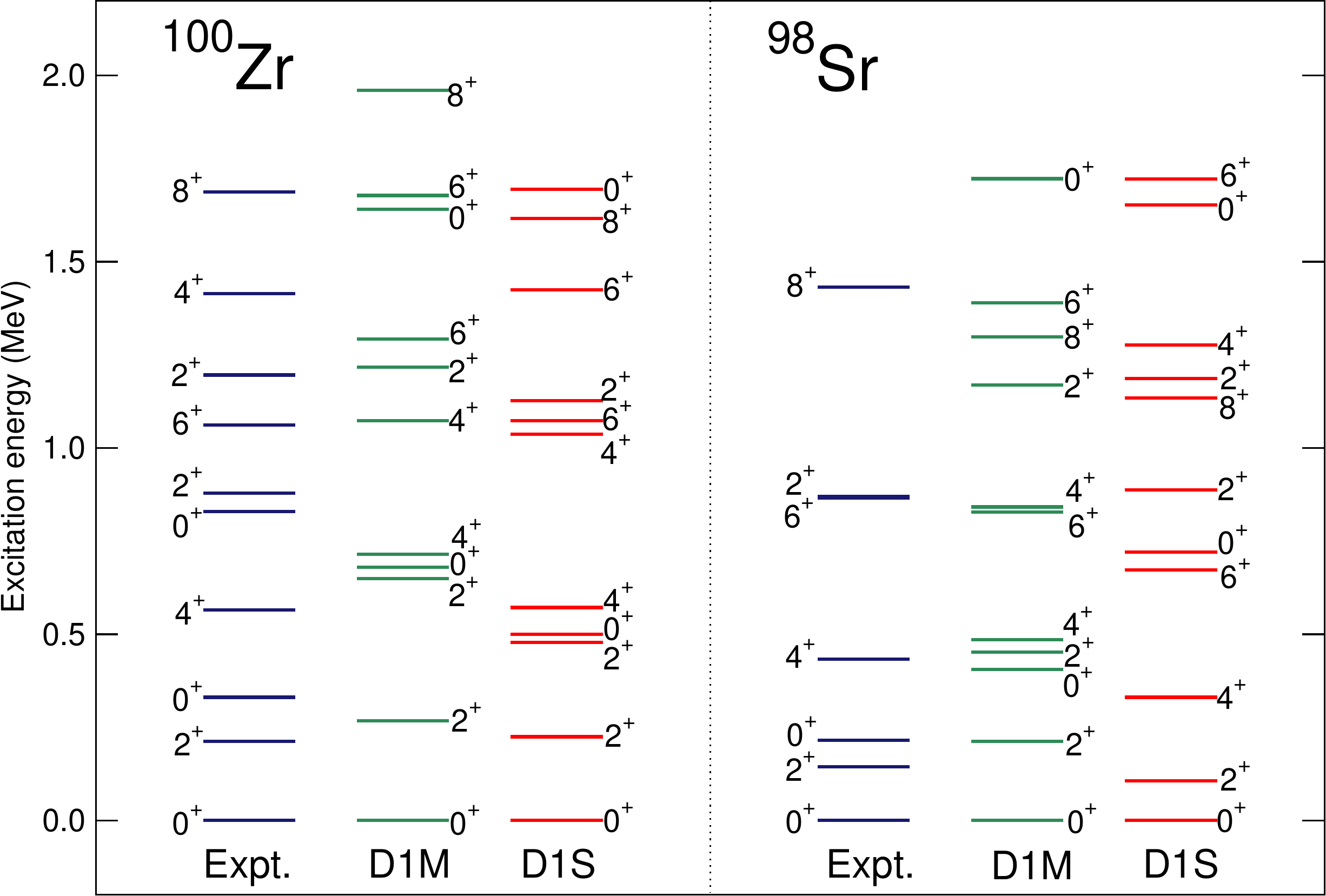}
\caption{(Color online) Comparison of the low-energy spectra for the
 $^{100}$Zr and $^{98}$Sr nuclei calculated with the parametrizations
 D1M and D1S of the Gogny-EDF. The experimental data
 are also shown for reference. }
\label{fig:d1sd1m_energy}
\end{center}
\end{figure*}

\begin{figure}[htb!]
\begin{center}
\includegraphics[width=0.8\linewidth]{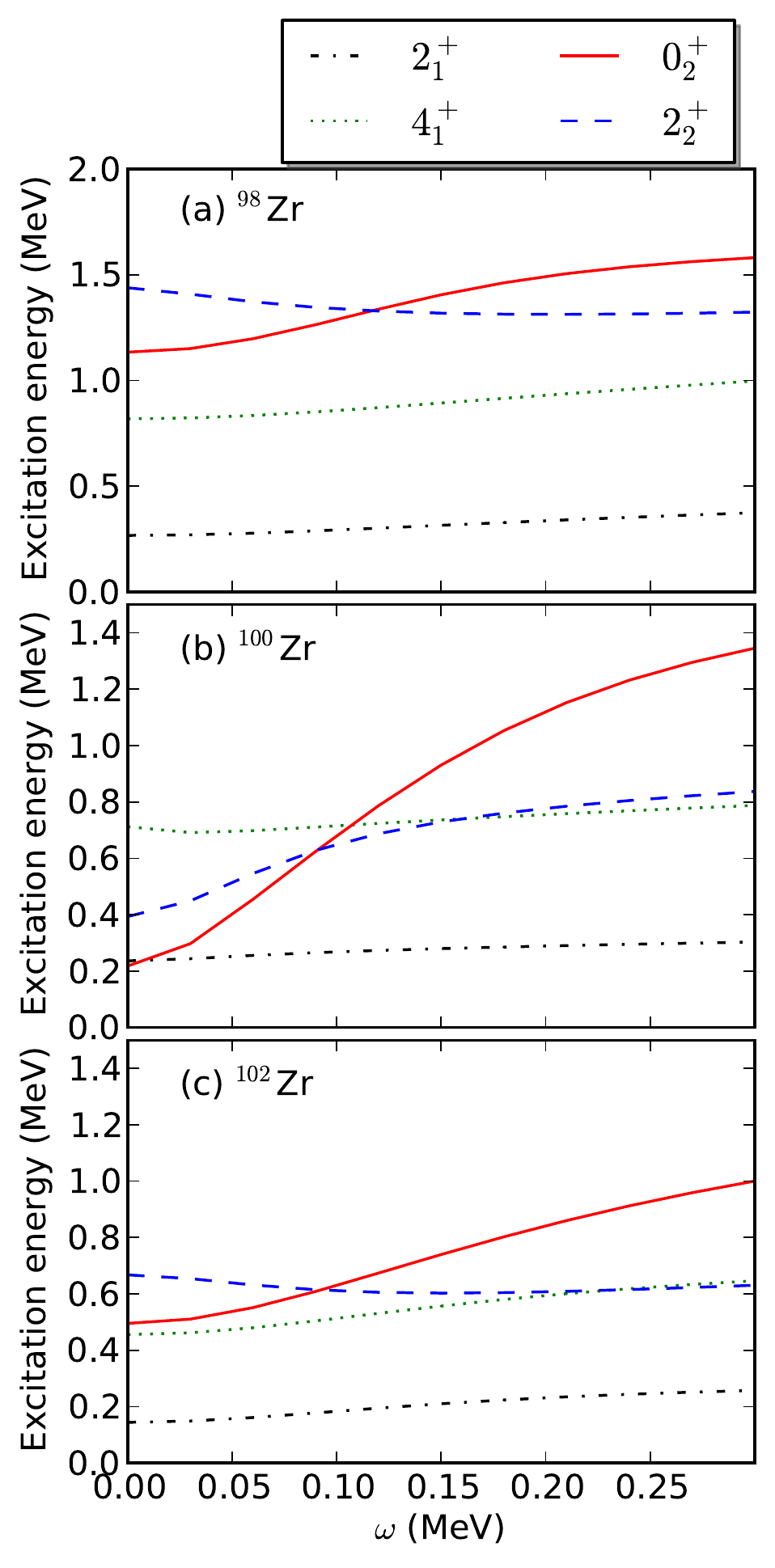}
\caption{(Color online) Excitation energies of the
$2^+_1$, $4^+_1$, $0^+_2$ and $2^+_2$ 
states of $^{98}$Zr (a), $^{100}$Zr (b) and $^{102}$Zr (c), as functions
of the mixing interaction strength $\omega$. Calculations are based on the 
parametrization D1M of the Gogny-EDF}
\label{fig:eomg}
\end{center}
\end{figure}

As pointed out in previous sections, there are two major factors
which could affect the description of the energy spectra, especially,
those of the excited $0^+$ states: 
one is the choice of the particular version of the EDF, and the other is
the choice of the mixing interaction strength $\omega$. 
In this subsection, we check the sensitivity of our results to these two factors. 

First, we show in Figs.~\ref{fig:d1sd1m_beta} and \ref{fig:d1sd1m_gamma}
the SCMF deformation energy curves for
the $N=60$ isotones [$^{104}$Ru (a), $^{102}$Mo (b), $^{100}$Zr (c) and 
 $^{98}$Sr (d)] calculated with
the D1S and D1M parametrizations of the Gogny-EDF, as 
functions of the axial deformation parameter $\beta$ (with
$\gamma=0^{\circ}$) and the non-axial deformation parameter $\gamma$
(with $\beta$ corresponding to the minimum at each $\gamma$ value),
respectively. 
In $^{104}$Ru [panel (a)] and
$^{102}$Mo [panel (b)], we do not observe striking differences in the topology of the
SCMF energy surfaces computed with the D1S and D1M parametrizations. 
In $^{100}$Zr and $^{98}$Sr, on the contrary, there are notable differences, especially, in the
energies of prolate and oblate minima and in the $\gamma$ softness. 

Regarding the $^{100}$Zr and $^{98}$Sr nuclei, we compare in Fig.~\ref{fig:d1sd1m_energy} the
energy spectra obtained with the parameters deduced from the D1S and D1M energy surfaces.  
In Fig.~\ref{fig:d1sd1m_energy}, the results based on the D1S
interaction suggest that, in both $^{100}$Zr and $^{98}$Sr, the energy levels for 
the yrast states, which are mainly composed of the oblate global minimum
(see also, Fig.~\ref{fig:frac}), are
more compressed than those based on the D1M interaction. 
The excitation energies of the non-yrast states, e.g., the $0^+_2$ and
$2^+_2$ states, which are mainly coming from the second lowest prolate mean-field
minimum, are rather dependent on the choice of the EDF. 
This is corroborated by the SCMF results shown in panel (c) [(d)] 
of Figs.~\ref{fig:d1sd1m_beta} and \ref{fig:d1sd1m_gamma}, where one sees 
that the energy difference between the prolate and oblate mean-field minima
is in the case of the D1M force larger (smaller) than in the case of the D1S
force.

Second, we display in Fig.~\ref{fig:eomg} the excitation energy of the
$2^+_1$, $4^+_1$, $0^+_2$ and $2^+_2$ 
states, calculated within the configuration mixing IBM, as a function
of the mixing interaction strength $\omega$, for $^{98}$Zr (a),
$^{100}$Zr (b) and $^{102}$Zr (c). Calculations are based on the 
parametrization D1M of the Gogny-EDF.
 The energies of the yrast ($2^+_1$ and $4^+_1$) states stay almost
constant with $\omega$, 
whereas those of the non-yrast ($0^+_2$ and $2^+_2$) states are more
sensitive to $\omega$. 
For $^{98,100}$Zr, the chosen value $\omega=0.1$ MeV  seems to be too
large to explain the experimental $0^+_2$ level energies of 
854 keV and 331 keV, respectively \cite{data}. 
For $^{102}$Zr, on the other hand, much larger value of the strength
$\omega$ could be required to account for the experimental $0^+_2$
energy of 895 keV.


\section{Concluding remarks\label{sec:summary}}

In this work, we have studied the shape evolution and coexistence in 
the neutron-rich nuclei $^{98-114}$Ru, $^{96-112}$Mo, $^{94-110}$Zr and 
$^{92-108}$Sr. We have resorted to the SCMF-to-IBM mapping procedure 
based on the Gogny-D1M EDF. The IBM parameters derived from such a 
procedure have  been used to compute the spectroscopic properties of 
the considered nuclei. In order to keep our analysis as simple as 
possible several approximations have been made. Our method describes 
reasonably well the evolution of the low-lying yrast and non-yrast 
states. Our results for  Ru and Mo nuclei suggest many $\gamma$-soft 
examples while some vivid examples of coexistence between strongly 
deformed prolate and weakly deformed oblate shapes have been found for 
the Zr and Sr nuclei.

Our calculations describe well the rapid structural change between 
$N=58$ and 60 in Zr and Sr nuclei. The analysis of the  Gogny-D1M  and 
mapped  IBM energy surfaces as well as the wave functions of the 
$0^+_1$ and $0^+_2$ states reveals that the sudden lowering of the 
energy  levels from $N=58$ to 60 in those nuclei is the consequence of 
the onset of large prolate deformations. 
From Fig.~\ref{fig:frac}, many of the Zr and Sr nuclei 
from $N=60$ till around the neutron mid-shell $N\approx 66$ exhibit a 
prolate ground state while their $0^+_2$ states are dominated by the  
oblate configuration. On the other hand, an oblate ground state is 
found for the heavier isotopes near $N=70$.

We have also pointed out several discrepancies between our predictions 
and the available experimental data. In particular, for many of the 
considered nuclei, the $0^+_2$ excited state is predicted to be too 
high. In Mo isotopes, for example, the $0^+_2$ energy level is 
systematically overestimated since the mixing is not introduced in most 
of the isotopes. The $0^+_2$ excitation energy is neither well 
described at $N=60$ and 62 in the case of Zr and Sr chains. This 
discrepancy could be related to the particular version of the Gogny-EDF 
employed in our calculations. In the case of $^{98}$Mo, for example, 
the SCMF energy surface displays only one minimum whereas previous 
Skyrme-HF+BCS calculations \cite{thomas2013} have found two minima. 
However, a second source for the discrepancy could also be the 
assumptions made at the IBM level. For example, the simplified form of 
the unperturbed Hamiltonian and a constant mixing strength may not be 
realistic enough for all studied nuclei. One could also use a  
different boson model space. Within this context, some refinement is 
still required to better constrain the IBM Hamiltonian. For example, 
both particle and hole pairs have been mapped onto the same boson 
image. However, a more realistic formulation would consider a single 
boson Hamiltonian where both particle-like and hole-like bosons are 
taken into account, rather than invoking several different unperturbed 
Hamiltonians. Nevertheless, we stress that the considered mapping 
procedure allows a systematic and computationally feasible description 
of medium-mass and heavy nuclei with several coexisting shapes. The 
method can also be used  to predict the spectroscopic properties of 
unexplored nuclei.

\begin{acknowledgments}

Author K.N. thanks financial support by the Japan 
Society of Promotion of Science. The  work of LMR was supported by the 
Spanish Ministerio de Econom\'ia y Competitividad (MINECO), under 
Contracts Nos. FIS2012-34479, FPA2015-65929 and FIS2015-63770.
\end{acknowledgments}

\bibliography{refs}
\end{document}